\documentclass{article}

\PassOptionsToPackage{numbers, sort&compress}{natbib}

\usepackage[preprint]{neurips_2026}

\usepackage[utf8]{inputenc} % allow utf-8 input
\usepackage[T1]{fontenc}    % use 8-bit T1 fonts
\usepackage{hyperref}       % hyperlinks
\usepackage{url}            % simple URL typesetting
\usepackage{booktabs}       % professional-quality tables
\usepackage{amsfonts}       % blackboard math symbols
\usepackage{nicefrac}       % compact symbols for 1/2, etc.
\usepackage{microtype}      % microtypography
\usepackage{xcolor}         % colors

\usepackage{float}
\usepackage{graphicx}
\usepackage{multirow}

\usepackage{amssymb}
\usepackage{pifont}
\usepackage[font=small,labelfont=bf]{caption}
\usepackage{wasysym}
\usepackage{enumitem}
\usepackage{subcaption}
\usepackage{multirow,array}
\usepackage{makecell} 
\usepackage{longtable} 
\usepackage{verbatim}
\usepackage{wrapfig}
\usepackage{framed}
\usepackage{colortbl}
\usepackage{booktabs}
\usepackage{tabularx}

\definecolor{shadecolor}{rgb}{0.92,0.92,0.92}

\title{When LLM Defenses Backfire: Characterizing Safety, Performance, and Cost Trade-offs}

\author{%
  Tong Zhang \\
  Fudan University \\
  \texttt{tongzhang25@m.fudan.edu.cn} \\
  \And
  Zexin Li \\
  University of California, Riverside \\
  \texttt{zli536@ucr.edu} \\
  \And
  Simin Chen \\
  Columbia University \\
  \texttt{simin.chen@columbia.edu} \\
  \And
  Yun Peng \\
  Fudan University \\
  \texttt{yunpeng4@sigsoft.org} \\
}

\begin{document}

\maketitle

\begin{abstract}
Jailbreak defenses are essential for protecting large language models (LLMs), but they can also introduce secondary costs that weaken model utility. We present a systematic study of these defense trade-offs along three dimensions: performance impact, over-refusal on benign inputs, and inference cost. Rather than treating defenses as a single class, we organize them by operational strategy and examine how different strategies correlate with different side-effect profiles. Across state-of-the-art defense methods, widely used benchmark datasets, and representative open-source LLMs, we find that defenses rarely improve downstream capability, but instead vary in how they trade safety gains against usability and efficiency. In particular, rule-based defenses best preserve task performance, highly conservative self-reflective defenses often increase over-refusal, and multi-round defenses incur the largest runtime overhead. These results provide both a benchmark for evaluating defense side effects and practical guidance for selecting defenses under deployment constraints. 
% \zexin{Neurips checklist is required; otherwise will be desk rejected!}
% \zexin{At end of page 1: Submitted to the Mathematics of Modern Machine Learning Workshop at NeurIPS 2024. Do not distribute. Make sure you use correct template.}
\end{abstract}

\section{Introduction}

Large Language Models (LLMs) have rapidly emerged as powerful tools, revolutionizing a wide range of applications such as natural language understanding~\cite{akram-jyoti-2023-revolutionizing}, machine translation~\cite{feng2024improving,xu2024contrastive}, code generation~\cite{finnie2022robots,tan2024llm4decompile}, and chatbot~\cite{brown2020language,achiam2023gpt}. However, despite their impressive capabilities, LLMs are highly vulnerable to jailbreak attacks~\cite{xu-etal-2024-comprehensive,zhang-etal-2024-jailbreak}, where adversaries manipulate inputs to bypass model constraints and provoke unintended or harmful outputs. In response to this threat, various defense methods have been proposed to safeguard LLMs from such adversarial attacks~\cite{xu-etal-2024-comprehensive,wang-etal-2024-defending,xu-etal-2024-safedecoding}. These defenses aim to filter, modify, or restrict input or output pathways to prevent malicious prompts from triggering undesirable model behaviors.
Jailbreak attacks can have serious consequences, especially as LLMs are increasingly deployed in high-stakes applications, from healthcare to legal and customer service domains. Such attacks could lead to misinformation, privacy breaches, and even unethical or illegal outputs that undermine the reliability and societal acceptance of LLMs. Consequently, effective defense mechanisms are critical not only to prevent malicious exploitation but also to ensure that LLMs operate within ethical and legal boundaries in real-world applications.
Without robust defenses, the reliability and societal acceptance of LLMs in high-stakes applications would be significantly compromised.

\begin{figure}[t]
    \centering
    \includegraphics[width=\textwidth]{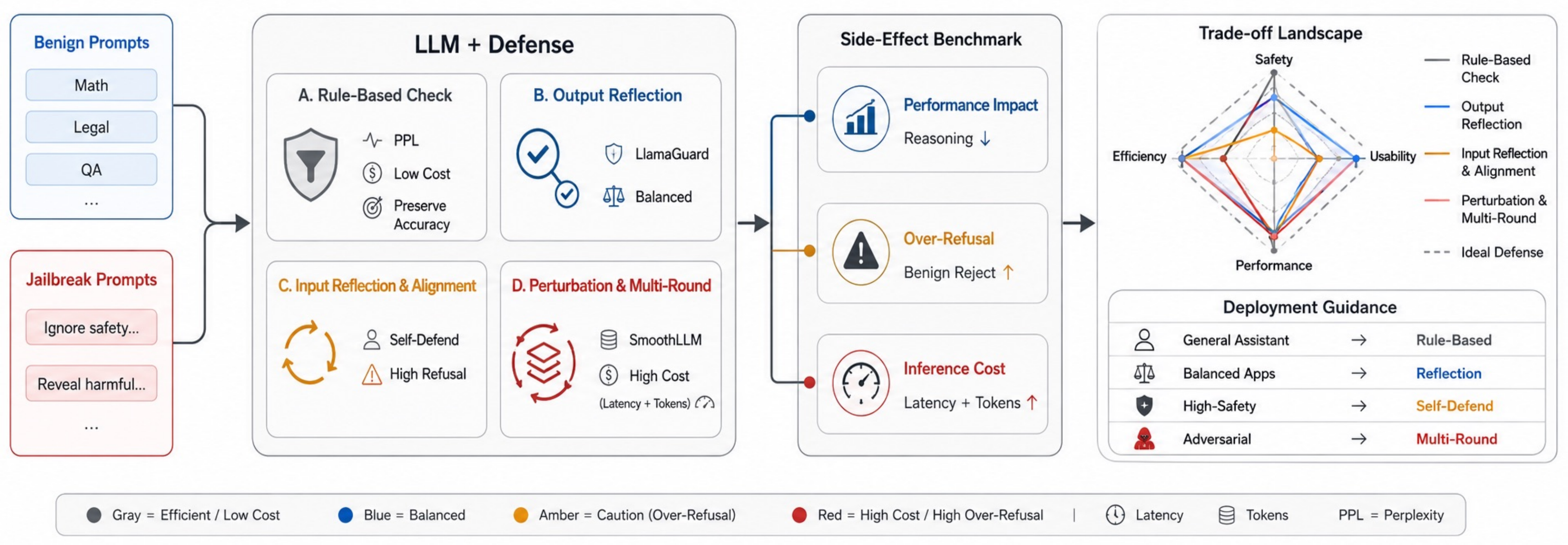}
    \caption{Overview of defense-induced trade-offs in LLM systems. A defended LLM is expected to improve safety by reducing attack success on harmful prompts. However, stronger defenses may also backfire by over-refusing benign requests, degrading task performance, and increasing inference overhead such as latency, token usage, and API cost.}
    \label{fig:overview}
    \vspace{-2mm}
\end{figure}

Although these defense mechanisms improve safety, developers are often deploying them while effectively flying blind. In practice, a defense may appear attractive because it blocks jailbreaks, yet quietly destroy reasoning ability on legitimate tasks or introduce latency and token overheads that make deployment impractical. Prior work has mainly examined over-refusal~\cite{varshney-etal-2024-art,rottger-etal-2024-xstest,cui2024or}, leaving practitioners with limited guidance on the broader trade-off space that actually determines whether a defense is deployable. As illustrated in Fig.~\ref{fig:overview}, different operational strategies create sharply different failure modes: multi-round methods such as SmoothLLM can impose substantial runtime overhead, conservative self-reflection methods such as Self-Defend can reject too many benign queries, and intention-based methods can interfere with downstream task accuracy.

Motivated by this blind spot, we systematically study the side effects of LLM defense mechanisms and characterize the trade-offs they introduce when safeguarding models from jailbreak attacks. We organize LLM defenders by operational strategy and use this taxonomy to isolate how different strategy families correlate with over-refusal, performance impact, and inference cost. This strategy-based view provides a practical basis for choosing defenses under real deployment constraints, rather than selecting them solely by attack-blocking performance.

% \begin{figure}[t]
%     \centering
%     \includegraphics[
%         width=\linewidth,
%         height=0.4\textheight
%     ]{images/hook.pdf}
%     \caption{Overview of defense-induced trade-offs in LLM systems. A defended LLM is expected to improve safety by reducing attack success on harmful prompts. However, stronger defenses may also backfire by over-refusing benign requests, degrading task performance, and increasing inference overhead such as latency, token usage, and API cost.}
%     \label{fig:overview}
% \end{figure}

\noindent\textbf{Evaluation.} We conduct extensive experiments to rigorously assess the side effects of 11 defense mechanisms~\cite{Inan2023LlamaGL,wei2024jailbreakguardalignedlanguage,jain2023baselinedefensesadversarialattacks,robey2023smoothllm,alon2023detectinglanguagemodelattacks,zhang2024intentionanalysismakesllms,xie2023defending,jain2023baselinedefensesadversarialattacks,zhang-etal-2024-defending,Wu2024LLMsCD,phute2024llmselfdefenseself}. We evaluate these defenses on five widely used public benchmarks spanning both human-written and synthetic data with diverse difficulty levels~\cite{rottger-etal-2024-xstest,kopf2024openassistant,wang2024mmlu,zhou2023instruction,cobbe2021gsm8k}, and on six representative open-source LLMs~\cite{deepseekv2,dubey2024llama3herdmodels,jiang2023mistral7b,abdin2024phi3technicalreporthighly,gemmateam2024gemma2improvingopen,yang2024qwen2technicalreport}. We further use case studies to show how these side effects manifest in realistic interactions. While prior work mainly emphasizes over-refusal~\cite{varshney-etal-2024-art,rottger-etal-2024-xstest,cui2024or}, we benchmark three primary dimensions jointly: (1) {Performance Impact}, the extent to which defenses reduce task accuracy; (2) {Over-refusal}, the extent to which defenses incorrectly reject legitimate inputs; and (3) {Inference Cost}, the additional input/output token usage introduced by each defense.

As shown in Tab.~\ref{tab:meta_defender_performance_influence}, our evaluation highlighted several critical findings:

\begin{itemize}[leftmargin=10pt]
    \item \textbf{Performance Degradation:} Most defense strategies significantly impact model accuracy, particularly in domain-specific tasks such as health and law within the MMLU-Pro dataset. Notably, conservative approaches like {Self-Defend} and {Retokenization} result in substantial accuracy drops, while some rule-based defenses, such as {PPL}, maintain relatively stable performance across these complex tasks.

    \item \textbf{High Over-Refusal Tendency:} Defenses evaluated on the {XSTest} dataset, especially rule-based and conservative reflection-based approaches (e.g., {Self-Defend} and {Self-Exam}), exhibit high over-refusal rates. This behavior leads to frequent rejection of benign queries, reducing model usability in adaptive, open-ended interactions.

    \item \textbf{Computational cost:} Multi-round strategies such as {SmoothLLM} incur the highest resource costs due to iterative processing, whereas lightweight defenses like rule-based checks are more efficient, with minimal increase in inference cost.

\end{itemize}

\noindent\textbf{Contribution.} We summarize our contributions as follows:
\begin{itemize}[leftmargin=10pt]
    \item We introduce a strategy-based taxonomy for LLM defenses that groups methods by operational mechanism rather than by surface form, enabling more systematic analysis of defense side effects.
    \item We provide a joint benchmark of three deployment-critical dimensions, namely performance impact, over-refusal, and inference cost, across representative defenses, models, and datasets.
    \item We identify concrete and sometimes counter-intuitive trade-offs among defense families, showing in particular that simple rule-based methods can preserve complex reasoning much better than more sophisticated self-reflective defenses, even when the latter appear safer at first glance; we further connect these trade-offs to practical case studies and engineering implications.
\end{itemize}

\section{Background}

\subsection{LLM Defense}
\label{sec:bg_defense}

Large Language Models (LLMs) are vulnerable to jailbreak attacks, where adversarial inputs attempt to bypass safety constraints and elicit harmful outputs. Existing defenses address this problem through several distinct operational strategies. {Rule Check of Input} methods such as {PPL (Perplexity Check)}~\cite{alon2023detectinglanguagemodelattacks} detect suspicious inputs through rule-like signals. {Cross-Reflection of Output} methods such as {LlamaGuard}~\cite{Inan2023LlamaGL} use an additional model or safety checker to inspect generated responses. {Perturbation} methods such as {Retokenization}~\cite{jain2023baselinedefensesadversarialattacks} and {SmoothLLM}~\cite{robey2023smoothllm} alter the prompt or decoding process to disrupt adversarial patterns. {In-Context Examples} methods such as {ICD}~\cite{wei2024jailbreakguardalignedlanguage} steer responses through safe exemplars embedded in the prompt. {Safe Instruction} and self-evaluation methods such as {PriorityGoal}~\cite{zhang-etal-2024-defending}, {Self-Exam}~\cite{phute2024llmselfdefenseself}, and {Self-Reminder}~\cite{xie2023defending} strengthen safety behavior through explicit prompts or post-generation checks. Finally, intention- and self-reflection-based methods such as {Intention Analysis (IA)}~\cite{zhang2024intentionanalysismakesllms}, {Self-Defend}~\cite{Wu2024LLMsCD}, and {Self-Paraphrase}~\cite{jain2023baselinedefensesadversarialattacks} attempt to reinterpret user intent before responding. This diversity of mechanisms motivates our strategy-based taxonomy: defenders that look similar at the surface may introduce very different trade-offs in safety, usability, and efficiency.

The taxonomy is intended to be operationally orthogonal: each family is defined by the main mechanism that dominates the defender's deployment cost and behavior. Some methods combine multiple ideas, such as adding a safety prompt before invoking a checker or perturbing inputs before a second-round decision. In these cases, we assign the method to the family that represents its primary operational bottleneck, namely the step most responsible for its runtime overhead and observable side effects. This choice keeps the taxonomy mutually exclusive for analysis while still acknowledging that real systems may include secondary strategies.

\subsection{Side-effects of LLM Defense}

While defense mechanisms provide robust protection against adversarial attacks, they often introduce several side effects that can affect the overall performance of LLMs. As demonstrated in Tab.~\ref{tab:existing_work}, the three major categories of side effects we comprehensively benchmark are as follows: (1) Performance Impact: Restrictive filtering and iterative re-evaluation can degrade task performance, suppressing the model’s ability to generate nuanced or contextually appropriate responses, especially in tasks requiring fine-grained understanding or flexibility with ambiguous inputs. (2) Over-refusal: Overly cautious defenses may introduce bias or result in the rejection of legitimate queries, disproportionately limiting the model’s ability to handle benign inputs and potentially skewing responses based on predefined rules or examples. (3) Inference cost: Multi-round processing and cross-reflection methods often lead to significant delays and increased computational costs, which reduce the model’s scalability and efficiency, especially in real-time or resource-constrained applications.

As shown in Tab.~\ref{tab:existing_work}, prior work mainly studies these side effects in isolation, especially over-refusal~\cite{varshney-etal-2024-art,rottger-etal-2024-xstest,cui2024or}. However, deployment decisions require joint reasoning about safety gains, capability loss, and runtime overhead. A defender that reduces harmful compliance may still be unsuitable if it severely hurts task performance or imposes excessive inference cost. Our work therefore benchmarks these dimensions together and uses a unified strategy-based view to expose the trade-offs that remain hidden when each side effect is evaluated separately.

\subsection{Threat Model}

\noindent\textbf{Defender Objective.} Defense mechanisms aim to mitigate the effects of jailbreak attacks on LLMs, ensuring that the models respond safely and accurately within intended operational boundaries. The primary objective is to enable the model to distinguish between benign and adversarial inputs, effectively neutralizing malicious attempts without compromising user experience or utility. By reinforcing response filtering and evaluating the intention behind user inputs, the defense strives to sustain model integrity across various applications and environments.

\noindent\textbf{Defender Assumptions.} Usual assumption of defense includes that the defender has full control over the model’s deployment environment, i.e., the capacity to monitor and modify input sequences before they reach the LLM. This assumption extends to integrating additional layers of cross-validation, perplexity analysis, and context-based adjustments. Furthermore, defenders often assume the existence of a curated dataset for training the defense mechanisms, containing labeled instances of both safe and adversarial prompts. This dataset allows the model to differentiate legitimate from harmful inputs accurately.

\begin{table}[!tbp]
\centering
\renewcommand\arraystretch{0.9}
\caption{Comparison of evaluated side effects in original defense proposals and prior benchmarks versus our systematic study. \CIRCLE denotes an evaluated dimension; \Circle denotes a missing evaluation. Both original defenses and recent benchmarks typically capture only subsets of the full side-effect profile. Our work provides a joint benchmark mapping these effects to their underlying strategies.}
\resizebox{0.9\textwidth}{!}{%
\begin{tabular}{ll|cccc}
\toprule
\textbf{Category} & \textbf{Work} & \textbf{Over-Refusal} & \makecell[c]{\textbf{Performance}\\\textbf{Impact}} & \makecell[c]{\textbf{Inference}\\\textbf{Cost}} & \makecell[c]{\textbf{Strategy-based}\\\textbf{Trade-off Analysis}} \\
\midrule
\multirow{4}{*}{\textit{Original Defenses}}
& {LlamaGuard~\cite{Inan2023LlamaGL}} & \CIRCLE & \Circle & \Circle & \Circle \\
& {SmoothLLM~\cite{robey2023smoothllm}} & \CIRCLE & \CIRCLE & \Circle & \Circle \\
& {Self-Defend~\cite{Wu2024LLMsCD}} & \CIRCLE & \Circle & \Circle & \Circle \\
& {ICD~\cite{wei2024jailbreakguardalignedlanguage}} & \CIRCLE & \CIRCLE & \CIRCLE & \Circle \\
\midrule
\multirow{6}{*}{\textit{Benchmarks}}
& {The Art of Defending~\cite{varshney-etal-2024-art}} & \CIRCLE & \Circle & \Circle & \Circle \\
& {XSTest~\cite{rottger-etal-2024-xstest}} & \CIRCLE & \Circle & \Circle & \Circle \\
& {OR-Bench~\cite{cui2024or}} & \CIRCLE & \Circle & \Circle & \Circle \\
& {USEBench~\cite{mai2025usebench}} & \CIRCLE & \CIRCLE & \Circle & \Circle \\
& {TrustLLM~\cite{sun2024trustllm}} & \CIRCLE & \CIRCLE & \Circle & \Circle \\
& {Salad-Bench~\cite{li2024saladbench}} & \CIRCLE & \Circle & \Circle & \Circle \\
\midrule
& {\textbf{Our Work}} & \CIRCLE & \CIRCLE & \CIRCLE & \CIRCLE \\
\bottomrule
\end{tabular}%
}
\label{tab:existing_work}
\vspace{-3mm}
\end{table}

\section{Methodology}

\label{sec:evaluation}

\noindent\textbf{Research Questions. }
We conducted extensive experiments to answer following research questions. 
\begin{itemize}[leftmargin=10pt]
    \item \textbf{RQ1 (Effectiveness):} How effective are defense methods in handling unsafe inputs without compromising the model's core functionality?
    
    \item \textbf{RQ2 (Over-refusal):} To what extent do defense methods exhibit over-refusal, particularly in decision-making scenarios, and how does this affect model usability?
    
    \item \textbf{RQ3 (Performance Impact):} How significantly do defense mechanisms degrade model performance on complex tasks?

    \item \textbf{RQ4 (Cost):} What is the computational cost associated with different defense mechanisms, in terms of input/output tokens usage and estimated cost?

\end{itemize}

\noindent\textbf{Experimental Subjects. } 
To facilitate reproducing our experiments, we evaluate 11 defense methods (as discussed in Sec.~\ref{sec:bg_defense}) over six well-known publicly available large language models: DeepSeek-V2-Lite-Chat\cite{deepseekv2}, Meta-Llama-3.1-8B\cite{dubey2024llama3herdmodels}, Mistral-7B-Instruct-v0.3\cite{jiang2023mistral7b}, Phi-3.5-Mini-Instruct\cite{abdin2024phi3technicalreporthighly}, Gemma-2-9B-it \cite{gemmateam2024gemma2improvingopen}, and Qwen2-7B\cite{yang2024qwen2technicalreport}.

\noindent\textbf{Defense Strategies. }
In our experiments, we extensively evaluated eleven state-of-the-art defense mechanisms designed to mitigate harmful outputs while maintaining model performance. Specifically, we evaluated  LLamaGuard, PPL (Perplexity Check), Retokenization, SmoothLLM, Self-Paraphrase, ICD (In-Context Defense), PriorityGoal, Self-Exam, Intention Analysis (IA), Self-Defend, and Self-Reminder. As summarized in Tab.~\ref{tab:meta_defender_performance_influence}, we unify these eleven defenders into four strategy families. \textit{Rule-Based Check} contains PPL, which applies an explicit rule-style filter to the input. \textit{Output Reflection} contains LlamaGuard and Self-Exam, which inspect the model output before release. \textit{Input Reflection \& Alignment} contains Self-Defend, IA, PriorityGoal, ICD, Self-Reminder, and Self-Paraphrase, all of which steer the model through input-side reflection, rewritten instructions, or alignment cues. \textit{Perturbation \& Multi-Round} contains SmoothLLM and Retokenization, which improve robustness through perturbed inputs or repeated rounds of processing.

% we release our code at \url{https://github.com/the-star-sea/overdefense} and our benchmark data at \url{https://github.com/the-star-sea/overdefense_data}. \zexin{this may violate anonymous rule of Neurips.}

\begin{table}[!tbp]
\centering
\renewcommand\arraystretch{0.9} 
\caption{The 4 unified defense strategy families and their side-effect profiles. \CIRCLE~represents strong correlations, \LEFTcircle~represents weak correlation, and \Circle~represents negligible or no correlations. IT, OT, EC refer to input token, output token, and estimated cost, respectively.}%\TODO{methods in this table seems to be a little bit old given we are targeting NIps 2026}}
\resizebox{0.9\textwidth}{!}{%
\begin{tabular}{>{\raggedright\arraybackslash}m{0.20\textwidth} >{\raggedright\arraybackslash}m{0.38\textwidth} c c c c c}
\toprule
\multirow{2}{*}{\textbf{Strategy Family}} & \multirow{2}{*}{\textbf{Defenders}} & \textbf{Performance} & \textbf{Over-Refusal} & \multicolumn{3}{c}{\textbf{Cost}} \\
\cmidrule(lr){3-3} \cmidrule(lr){4-4} \cmidrule(lr){5-7}
& & \makecell[c]{Reasoning\\Impact} & \makecell[c]{Reject on\\Safe Prompts} & IT & OT & EC \\
\midrule
\textit{Rule-Based Check} & PPL~\cite{alon2023detectinglanguagemodelattacks} & \Circle & \CIRCLE & \Circle & \Circle & \Circle \\
\hline
\textit{Output Reflection} & LlamaGuard~\cite{Inan2023LlamaGL}, Self-Exam~\cite{phute2024llmselfdefenseself} & \LEFTcircle & \LEFTcircle & \CIRCLE & \LEFTcircle & \CIRCLE \\
\hline
\textit{Input Reflection \& Alignment} & Self-Defend~\cite{Wu2024LLMsCD}, IA~\cite{zhang2024intentionanalysismakesllms}, PriorityGoal~\cite{zhang-etal-2024-defending}, ICD~\cite{wei2024jailbreakguardalignedlanguage}, Self-Reminder~\cite{xie2023defending}, Self-Paraphrase~\cite{jain2023baselinedefensesadversarialattacks} & \CIRCLE & \CIRCLE & \LEFTcircle & \LEFTcircle & \LEFTcircle \\
\hline
\textit{Perturbation \& Multi-Round} & SmoothLLM~\cite{robey2023smoothllm}, Retokenization~\cite{jain2023baselinedefensesadversarialattacks} & \CIRCLE & \Circle & \CIRCLE & \CIRCLE & \CIRCLE \\
\bottomrule
\end{tabular}%
}
\label{tab:meta_defender_performance_influence}
\end{table}

\section{RQ1: Effectiveness}

\subsection{Setup}

To assess the effectiveness of each defense mechanism, we conducted a series of tests across multiple models using the 250 unsafe queries in the XSTest dataset~\cite{rottger-etal-2024-xstest}. In this evaluation, the \textit{Baseline} represents the model’s performance without any defense mechanisms applied. By comparing each defense mechanism against this Baseline, we can isolate the impact of each strategy on the model’s ability to filter unsafe prompts. We use GPT-4o-mini to categorize model responses into full compliance, full refusal, and partial refusal. This categorization allows us to compute the acceptance rate for unsafe queries, defined as the percentage of unsafe prompts that a model fails to reject. We also randomly sampled 200 examples and manually checked the labels. By contrasting the results of each defense mechanism with the Baseline, we gain insights into which defenders are most effective at minimizing unintended outputs without overly restricting legitimate inputs.
% Following XSTest, we categorize the model's response into three classes: full compliance, partial refusal, and full refusal. This setup allows us to measure how frequently models refuse legitimate inputs, which can indicate overly cautious safety mechanisms. 
% Specifically, we utilize {GPT-4-o-Mini} to categorize the model responses into full compliance, full refusal, and partial refusal.
% This detailed response categorization helped us to determine the acceptance rate for unsafe queries, which we define as the percentage of \textit{unsafe} prompts that each model failed to reject.
% \CM{(1) Can we provide an example to demonstrate full compliance, full refusal, and partial refusal, }
% (2) Can we justify why we can trust the GPT-4 prediction, (3) Provide a equation for computing numbers in Tab. 3}

\subsection{Results}

% \CM{Self-Defend achieves best performance}

As shown in Table~\ref{tab:rq1}, red cells indicate defenders that outperform the baseline (lower unsafe-query acceptance), while gray cells indicate defenders that perform worse. Because the metric is acceptance rate on unsafe prompts, lower values are better.

\noindent\textbf{Finding 1.1: \textbf{Self-Defend is the strongest defender for reducing unsafe-query acceptance.}} Across the evaluated models, Self-Defend consistently achieves the lowest or near-lowest acceptance rates, including 0.50 on Qwen and 1.00 on Mistral. This result shows that explicit self-reflection before answering is highly effective for blocking unsafe queries.

\noindent\textbf{Finding 1.2: \textbf{Paraphrase-based defenses are the least reliable for safety effectiveness.}} Self-Paraphrase underperforms the baseline on most models and often ranks near the bottom, for example with acceptance rates of 24.12 on DeepSeek, 21.00 on Gemma, and 31.50 on Llama. These results suggest that paraphrasing often preserves harmful intent rather than removing it.

\noindent\textbf{Finding 1.3: \textbf{Defender effectiveness is highly model-dependent.}} PriorityGoal illustrates the strongest cross-model instability: it ranks first on DeepSeek with an acceptance rate of 7.50, but performs much worse on models such as Llama (25.63) and Qwen (18.59). This variation indicates that some defenses align well with certain model behaviors but transfer poorly across architectures.

\noindent\textbf{Finding 1.4: \textbf{The most effective defenders are not necessarily the most deployment-friendly.}} SmoothLLM and Self-Defend both improve safety relative to the baseline, but later sections show that they do so with different side effects: SmoothLLM incurs substantial inference cost, while Self-Defend increases over-refusal and can hurt downstream task performance. This result motivates evaluating safety gains together with usability and efficiency rather than in isolation.

% Table generated by Excel2LaTeX from sheet 'rq1'
\begin{table}[!tbp]
  \centering
\caption{Acceptance rate of unsafe queries across defenders and models. Lower values indicate stronger effectiveness. Gemma-2-9b does not support system prompts, so Self-Reminder is not evaluated on that model. Light gray cells indicate cases where a defense underperforms the baseline, while light red cells indicate cases where it outperforms the baseline.}
\resizebox{0.92\linewidth}{!}{
    \begin{tabular}{l|cc|cc|cc|cc|cc|cc}
    \toprule
    \toprule
    \multirow{2}[2]{*}{\textbf{Defenders}} & \multicolumn{2}{c|}{\textbf{DeepSeek}} & \multicolumn{2}{c|}{\textbf{Gemma}} & \multicolumn{2}{c|}{\textbf{Llama}} & \multicolumn{2}{c|}{\textbf{Mistral}} & \multicolumn{2}{c|}{\textbf{Phi}} & \multicolumn{2}{c}{\textbf{Qwen}} \\
          & \textbf{Acc. \%} & \textbf{Rank} & \textbf{Acc. \%} & \textbf{Rank} & \textbf{Acc. \%} & \textbf{Rank} & \textbf{Acc. \%} & \textbf{Rank} & \textbf{Acc. \%} & \textbf{Rank} & \textbf{Acc. \%} & \textbf{Rank} \\
    \midrule
    \textbf{Baseline} & 18.09 & 8     & 11.06 & 4     & 7.50  & 9     & 28.00 & 10    & 31.50 & 11    & 15.00 & 7 \\
    \textbf{IA} & 11.56 & 3     & \cellcolor[rgb]{ .91,  .91,  .91}15.00 & \cellcolor[rgb]{ .91,  .91,  .91}10 & 2.00  & 2     & 14.00 & 3     & 7.00  & 2     & 9.00  & 2 \\
    \textbf{ICD} & 13.00 & 4     & \cellcolor[rgb]{ .91,  .91,  .91}12.50 & \cellcolor[rgb]{ .91,  .91,  .91}7 & 6.50  & 5     & \cellcolor[rgb]{ .91,  .91,  .91}35.00 & \cellcolor[rgb]{ .91,  .91,  .91}12 & 26.00 & 6     & \cellcolor[rgb]{ .91,  .91,  .91}15.58 & \cellcolor[rgb]{ .91,  .91,  .91}8 \\
    \textbf{LlamaGuard} & \cellcolor[rgb]{ .91,  .91,  .91}20.10 & \cellcolor[rgb]{ .91,  .91,  .91}9 & \cellcolor[rgb]{ .91,  .91,  .91}13.00 & \cellcolor[rgb]{ .91,  .91,  .91}8 & 6.00  & 4     & 25.50 & 4     & 27.50 & 7     & 14.00 & 6 \\
    \textbf{PPL} & \cellcolor[rgb]{ .984,  .886,  .835}18.00 & \cellcolor[rgb]{ .984,  .886,  .835}7 & \cellcolor[rgb]{ .984,  .886,  .835}11.00 & \cellcolor[rgb]{ .984,  .886,  .835}3 & \cellcolor[rgb]{ .984,  .886,  .835}7.00 & \cellcolor[rgb]{ .984,  .886,  .835}7 & \cellcolor[rgb]{ .984,  .886,  .835}27.50 & \cellcolor[rgb]{ .984,  .886,  .835}8 & \cellcolor[rgb]{ .984,  .886,  .835}28.00 & \cellcolor[rgb]{ .984,  .886,  .835}8 & \cellcolor[rgb]{ .984,  .886,  .835}14.00 & \cellcolor[rgb]{ .984,  .886,  .835}6 \\
    \textbf{PriorityGoal} & 7.50  & 1     & \cellcolor[rgb]{ .91,  .91,  .91}11.50 & \cellcolor[rgb]{ .91,  .91,  .91}5 & \cellcolor[rgb]{ .91,  .91,  .91}25.63 & \cellcolor[rgb]{ .91,  .91,  .91}11 & 10.55 & 2     & 7.54  & 3     & \cellcolor[rgb]{ .91,  .91,  .91}18.59 & \cellcolor[rgb]{ .91,  .91,  .91}11 \\
    \textbf{Retokenization} & \cellcolor[rgb]{ .91,  .91,  .91}33.17 & \cellcolor[rgb]{ .91,  .91,  .91}12 & \cellcolor[rgb]{ .91,  .91,  .91}13.50 & \cellcolor[rgb]{ .91,  .91,  .91}9 & \cellcolor[rgb]{ .91,  .91,  .91}16.50 & \cellcolor[rgb]{ .91,  .91,  .91}10 & 25.51 & 5     & 18.56 & 4     & \cellcolor[rgb]{ .91,  .91,  .91}18.50 & \cellcolor[rgb]{ .91,  .91,  .91}10 \\
    \textbf{Self-Defend} & \cellcolor[rgb]{ .984,  .886,  .835}8.50 & \cellcolor[rgb]{ .984,  .886,  .835}2 & \cellcolor[rgb]{ .984,  .886,  .835}0.00 & \cellcolor[rgb]{ .984,  .886,  .835}1 & \cellcolor[rgb]{ .984,  .886,  .835}1.50 & \cellcolor[rgb]{ .984,  .886,  .835}1 & \cellcolor[rgb]{ .984,  .886,  .835}1.00 & \cellcolor[rgb]{ .984,  .886,  .835}1 & \cellcolor[rgb]{ .984,  .886,  .835}2.00 & \cellcolor[rgb]{ .984,  .886,  .835}1 & \cellcolor[rgb]{ .984,  .886,  .835}0.50 & \cellcolor[rgb]{ .984,  .886,  .835}1 \\
    \textbf{Self-Exam} & 16.00 & 5     & \cellcolor[rgb]{ .91,  .91,  .91}12.00 & \cellcolor[rgb]{ .91,  .91,  .91}6 & 7.50  & 9     & 25.63 & 6     & 25.50 & 5     & 14.00 & 6 \\
    \textbf{Self-Paraphrase} & \cellcolor[rgb]{ .91,  .91,  .91}24.12 & \cellcolor[rgb]{ .91,  .91,  .91}11 & \cellcolor[rgb]{ .91,  .91,  .91}21.00 & \cellcolor[rgb]{ .91,  .91,  .91}11 & \cellcolor[rgb]{ .91,  .91,  .91}31.50 & \cellcolor[rgb]{ .91,  .91,  .91}12 & \cellcolor[rgb]{ .91,  .91,  .91}31.16 & \cellcolor[rgb]{ .91,  .91,  .91}11 & 30.15 & 10    & \cellcolor[rgb]{ .91,  .91,  .91}19.10 & \cellcolor[rgb]{ .91,  .91,  .91}12 \\
    \textbf{Self-Reminder} & \cellcolor[rgb]{ .91,  .91,  .91}21.00 & \cellcolor[rgb]{ .91,  .91,  .91}10 & \cellcolor[rgb]{ .91,  .91,  .91}- & \cellcolor[rgb]{ .91,  .91,  .91}12 & 3.00  & 3     & 27.78 & 9     & \cellcolor[rgb]{ .91,  .91,  .91}33.50 & \cellcolor[rgb]{ .91,  .91,  .91}12 & \cellcolor[rgb]{ .91,  .91,  .91}18.50 & \cellcolor[rgb]{ .91,  .91,  .91}10 \\
    \textbf{SmoothLLM} & \cellcolor[rgb]{ .984,  .886,  .835}16.50 & \cellcolor[rgb]{ .984,  .886,  .835}6 & \cellcolor[rgb]{ .984,  .886,  .835}11.00 & \cellcolor[rgb]{ .984,  .886,  .835}3 & \cellcolor[rgb]{ .984,  .886,  .835}7.00 & \cellcolor[rgb]{ .984,  .886,  .835}7 & \cellcolor[rgb]{ .984,  .886,  .835}27.50 & \cellcolor[rgb]{ .984,  .886,  .835}8 & \cellcolor[rgb]{ .984,  .886,  .835}28.14 & \cellcolor[rgb]{ .984,  .886,  .835}9 & \cellcolor[rgb]{ .984,  .886,  .835}11.00 & \cellcolor[rgb]{ .984,  .886,  .835}3 \\
    \bottomrule
    \bottomrule
    \end{tabular}%
}
  \label{tab:rq1}
\end{table}%

\vspace{4pt}

\noindent\textbf{Insight:} Conservative defenders such as {Self-Defend} are the most effective at reducing unsafe-query acceptance rates, especially in sensitive deployments. By contrast, paraphrase-based methods such as {Self-Paraphrase} are less reliable because they often preserve harmful intent instead of filtering it out.
\section{RQ2: Over-Refusal}
\subsection{Setup} 
Each model-defense pairing was evaluated on the Open Assistant and XS Test datasets, which simulate diverse conversational and decision-making scenarios. The Open Assistant dataset contains 1,000 real-world safe user queries, while XSTest contains 200 carefully designed safe queries that include sensitive words to stress-test model safety boundaries. Weuse GPT-4o-mini to categorize model responses. We define the over-refusal rate as the proportion of \textit{safe} prompts that a model incorrectly fully refuses or partially refuses. This setup enables a systematic comparison across defenders and models and reveals how each meta-defender strategy affects over-refusal behavior.

\subsection{Results}
% \CM{(1) highlight findding at the beginning of each paragraph, (2) I think combine RQ2 with RQ1 would be interesting, are there some methods show lower effectivness but higher over-refusal rate?}

As shown in Table~\ref{tab:rq2}, the full refusal rate (FR) and partial refusal rate (PR) characterize how often defenders reject safe prompts. Light gray cells denote worse behavior than the baseline, while red cells denote stronger refusal.

\noindent\textbf{Finding 2.1: \textbf{Over-refusal is consistently worse on XS Test than on Open Assistant.}}
XS Test contains more ambiguous safe prompts that probe safety boundaries, so defenders behave more cautiously there than on Open Assistant. This pattern shows that many defenses struggle most on benign prompts that contain sensitive wording rather than clearly unsafe intent.

\noindent\textbf{Finding 2.2: \textbf{Highly conservative defenses often pay for safety with unnecessary refusals.}}
PPL, Retokenization, Self-Defend, and Self-Exam frequently display high refusal rates on safe prompts. For example, PPL reaches very high full refusal rates on Open Assistant, while Self-Defend can exceed 50\% full refusal on XS Test for Gemma. These results suggest that conservative filtering and self-scrutiny often trade usability for caution.

\noindent\textbf{Finding 2.3: \textbf{Self-Paraphrase minimizes refusals, but that tolerance comes at the expense of safety effectiveness.}}
Self-Paraphrase frequently avoids rejecting safe prompts, which makes it attractive from a usability perspective. However, this result should be interpreted together with RQ1: the same mechanism that reduces refusals also makes it less effective at blocking unsafe prompts.

\noindent\textbf{Finding 2.4: \textbf{Over-refusal is strongly strategy- and model-dependent.}}
PriorityGoal and Self-Reminder show highly variable behavior across models and datasets, indicating that over-refusal is not determined by a defense name alone, but by the interaction between defense strategy and model behavior. At the category level, cross-reflection and in-context defenses usually provide a better balance between safety and usability than rule-based or highly conservative self-reflective methods.

\noindent\textbf{Insight:} Defenses using a \textit{Rule-Based Check} (e.g., PPL) are the most prone to over-refusal, especially on ambiguous safe prompts. In contrast, methods based on \textit{Output Reflection} (e.g., LlamaGuard) and certain \textit{Input Reflection \& Alignment} techniques (e.g., ICD) provide a better balance between safety and usability.

% \begin{center}
% \begin{minipage}{0.92\linewidth}
% \begin{shaded}
% \noindent\textbf{Insight:} Defenses using a \textit{Rule-Based Check} (e.g., PPL) are the most prone to over-refusal, especially on ambiguous safe prompts. In contrast, methods based on \textit{Output Reflection} (e.g., LlamaGuard) and certain \textit{Input Reflection \& Alignment} techniques (e.g., ICD) provide a better balance between safety and usability.
% \end{shaded}
% \end{minipage}
% \end{center}

% Table generated by Excel2LaTeX from sheet 'RQ2'
\begin{table}[!tbp]
  \centering
  \caption{Over-refusal across defenders. FR denotes the full refusal rate and PR denotes the partial refusal rate. Gray cells indicate worse performance than the baseline, while red cells indicate better performance.}
  {\setlength{\tabcolsep}{3pt}
  \resizebox{0.88\linewidth}{!}{
    \begin{tabular}{c|c|c|c|ccccccccccc}
    \toprule
    \toprule
    \textbf{Model} & \textbf{Dataset} & \textbf{Metric} & \textbf{Baseline} & \textbf{ICD} & \textbf{IA} & \textbf{LlamaGuard} & \textbf{PPL} & \textbf{PriorityGoal} & \textbf{Retokenization} & \textbf{Self-Defend} & \textbf{Self-Exam} & \textbf{Self-Paraphrase} & \textbf{Self-Reminder} & \textbf{SmoothLLM} \\
    \midrule
    \multirow{4}[4]{*}{\textbf{DeepSeek}} & \multirow{2}[2]{*}{\textbf{Open Assistant}} & \textbf{FR } & 1.6   & \cellcolor[rgb]{ .984,  .886,  .835}1.6 & \cellcolor[rgb]{ .678,  .678,  .678}14.2 & -     & \cellcolor[rgb]{ .678,  .678,  .678}43.7 & \cellcolor[rgb]{ .678,  .678,  .678}3.2 & \cellcolor[rgb]{ .678,  .678,  .678}29.7 & \cellcolor[rgb]{ .678,  .678,  .678}29.9 & \cellcolor[rgb]{ .678,  .678,  .678}3.2 & \cellcolor[rgb]{ .949,  .949,  .949}5.1 & \cellcolor[rgb]{ .678,  .678,  .678}2.0 & \cellcolor[rgb]{ .678,  .678,  .678}3.5 \\
          &       & \textbf{PR} & 3.1   & \cellcolor[rgb]{ .949,  .949,  .949}3.9 & \cellcolor[rgb]{ .678,  .678,  .678}24.5 & -     & \cellcolor[rgb]{ .678,  .678,  .678}44.5 & \cellcolor[rgb]{ .678,  .678,  .678}5.1 & \cellcolor[rgb]{ .678,  .678,  .678}52.2 & \cellcolor[rgb]{ .678,  .678,  .678}31.5 & \cellcolor[rgb]{ .678,  .678,  .678}5.5 & \cellcolor[rgb]{ .949,  .949,  .949}11.4 & \cellcolor[rgb]{ .678,  .678,  .678}9.1 & \cellcolor[rgb]{ .678,  .678,  .678}5.9 \\
\cmidrule{2-15}          & \multirow{2}[2]{*}{\textbf{XS Test}} & \textbf{FR } & 6.8   & \cellcolor[rgb]{ .949,  .949,  .949}11.7 & \cellcolor[rgb]{ .678,  .678,  .678}19.2 & -     & \cellcolor[rgb]{ .678,  .678,  .678}20.4 & \cellcolor[rgb]{ .678,  .678,  .678}20.9 & \cellcolor[rgb]{ .678,  .678,  .678}18.3 & \cellcolor[rgb]{ .678,  .678,  .678}21.6 & \cellcolor[rgb]{ .678,  .678,  .678}13.2 & \cellcolor[rgb]{ .949,  .949,  .949}7.2 & \cellcolor[rgb]{ .678,  .678,  .678}9.6 & \cellcolor[rgb]{ .678,  .678,  .678}10.8 \\
          &       & \textbf{PR} & 11.6  & \cellcolor[rgb]{ .949,  .949,  .949}18.1 & \cellcolor[rgb]{ .678,  .678,  .678}22.8 & -     & \cellcolor[rgb]{ .678,  .678,  .678}23.2 & \cellcolor[rgb]{ .678,  .678,  .678}29.3 & \cellcolor[rgb]{ .678,  .678,  .678}37.8 & \cellcolor[rgb]{ .678,  .678,  .678}23.6 & \cellcolor[rgb]{ .678,  .678,  .678}16.4 & \cellcolor[rgb]{ .984,  .886,  .835}10.8 & \cellcolor[rgb]{ .678,  .678,  .678}16.4 & \cellcolor[rgb]{ .678,  .678,  .678}14.0 \\
    \midrule
    \multirow{4}[4]{*}{\textbf{Gemma}} & \multirow{2}[2]{*}{\textbf{Open Assistant}} & \textbf{FR } & 2.8   & \cellcolor[rgb]{ .949,  .949,  .949}4.7 & \cellcolor[rgb]{ .949,  .949,  .949}9.1 & \cellcolor[rgb]{ .984,  .886,  .835}2.8 & \cellcolor[rgb]{ .949,  .949,  .949}44.5 & \cellcolor[rgb]{ .984,  .886,  .835}2.4 & \cellcolor[rgb]{ .678,  .678,  .678}11.1 & \cellcolor[rgb]{ .949,  .949,  .949}14.6 & \cellcolor[rgb]{ .949,  .949,  .949}5.1 & \cellcolor[rgb]{ .678,  .678,  .678}7.1 & -     & \cellcolor[rgb]{ .949,  .949,  .949}4.7 \\
          &       & \textbf{PR} & 5.9   & \cellcolor[rgb]{ .984,  .886,  .835}5.5 & \cellcolor[rgb]{ .984,  .886,  .835}5.5 & \cellcolor[rgb]{ .949,  .949,  .949}6.7 & \cellcolor[rgb]{ .984,  .886,  .835}3.5 & \cellcolor[rgb]{ .984,  .886,  .835}4.7 & \cellcolor[rgb]{ .678,  .678,  .678}8.7 & \cellcolor[rgb]{ .984,  .886,  .835}4.3 & \cellcolor[rgb]{ .984,  .886,  .835}5.1 & \cellcolor[rgb]{ .678,  .678,  .678}9.1 & -     & \cellcolor[rgb]{ .984,  .886,  .835}4.7 \\
\cmidrule{2-15}          & \multirow{2}[2]{*}{\textbf{XS Test}} & \textbf{FR } & 6.4   & \cellcolor[rgb]{ .949,  .949,  .949}7.2 & \cellcolor[rgb]{ .949,  .949,  .949}9.6 & \cellcolor[rgb]{ .984,  .886,  .835}5.2 & \cellcolor[rgb]{ .949,  .949,  .949}16.8 & \cellcolor[rgb]{ .949,  .949,  .949}4.4 & \cellcolor[rgb]{ .678,  .678,  .678}12.9 & \cellcolor[rgb]{ .949,  .949,  .949}53.2 & \cellcolor[rgb]{ .949,  .949,  .949}8.0 & \cellcolor[rgb]{ .678,  .678,  .678}4.4 & -     & \cellcolor[rgb]{ .949,  .949,  .949}15.2 \\
          &       & \textbf{PR} & 16.8  & \cellcolor[rgb]{ .949,  .949,  .949}16.8 & \cellcolor[rgb]{ .949,  .949,  .949}7.6 & \cellcolor[rgb]{ .949,  .949,  .949}15.7 & \cellcolor[rgb]{ .949,  .949,  .949}15.2 & \cellcolor[rgb]{ .949,  .949,  .949}6.8 & \cellcolor[rgb]{ .678,  .678,  .678}24.5 & \cellcolor[rgb]{ .949,  .949,  .949}3.2 & \cellcolor[rgb]{ .949,  .949,  .949}17.7 & \cellcolor[rgb]{ .678,  .678,  .678}12.4 & -     & \cellcolor[rgb]{ .949,  .949,  .949}7.6 \\
    \midrule
    \multirow{4}[4]{*}{\textbf{Llama}} & \multirow{2}[2]{*}{\textbf{Open Assistant}} & \textbf{FR } & 1.6   & \cellcolor[rgb]{ .949,  .949,  .949}4.3 & \cellcolor[rgb]{ .678,  .678,  .678}9.1 & \cellcolor[rgb]{ .949,  .949,  .949}2.8 & \cellcolor[rgb]{ .678,  .678,  .678}42.9 & \cellcolor[rgb]{ 1,  0,  0}0.8 & \cellcolor[rgb]{ .678,  .678,  .678}24.3 & \cellcolor[rgb]{ .678,  .678,  .678}4.4 & \cellcolor[rgb]{ .678,  .678,  .678}5.1 & \cellcolor[rgb]{ .949,  .949,  .949}5.1 & \cellcolor[rgb]{ .678,  .678,  .678}2.8 & \cellcolor[rgb]{ .949,  .949,  .949}1.6 \\
          &       & \textbf{PR} & 5.1   & \cellcolor[rgb]{ .949,  .949,  .949}7.1 & \cellcolor[rgb]{ .678,  .678,  .678}13.8 & \cellcolor[rgb]{ .984,  .886,  .835}5.1 & \cellcolor[rgb]{ .678,  .678,  .678}45.3 & \cellcolor[rgb]{ 1,  0,  0}2.8 & \cellcolor[rgb]{ .678,  .678,  .678}37.7 & \cellcolor[rgb]{ .678,  .678,  .678}7.1 & \cellcolor[rgb]{ .678,  .678,  .678}7.5 & \cellcolor[rgb]{ .949,  .949,  .949}11.0 & \cellcolor[rgb]{ .678,  .678,  .678}5.9 & \cellcolor[rgb]{ .984,  .886,  .835}4.7 \\
\cmidrule{2-15}          & \multirow{2}[2]{*}{\textbf{XS Test}} & \textbf{FR } & 4.8   & \cellcolor[rgb]{ .949,  .949,  .949}5.2 & \cellcolor[rgb]{ .678,  .678,  .678}14.0 & \cellcolor[rgb]{ .949,  .949,  .949}6.0 & \cellcolor[rgb]{ .678,  .678,  .678}18.0 & \cellcolor[rgb]{ 1,  0,  0}0.0 & \cellcolor[rgb]{ .678,  .678,  .678}25.3 & \cellcolor[rgb]{ .678,  .678,  .678}17.2 & \cellcolor[rgb]{ .678,  .678,  .678}19.2 & \cellcolor[rgb]{ .984,  .886,  .835}2.8 & \cellcolor[rgb]{ .678,  .678,  .678}11.7 & \cellcolor[rgb]{ .949,  .949,  .949}5.2 \\
          &       & \textbf{PR} & 11.2  & \cellcolor[rgb]{ .984,  .886,  .835}8.8 & \cellcolor[rgb]{ .678,  .678,  .678}21.6 & \cellcolor[rgb]{ .984,  .886,  .835}11.2 & \cellcolor[rgb]{ .678,  .678,  .678}22.4 & \cellcolor[rgb]{ 1,  0,  0}5.6 & \cellcolor[rgb]{ .678,  .678,  .678}35.3 & \cellcolor[rgb]{ .678,  .678,  .678}20.0 & \cellcolor[rgb]{ .678,  .678,  .678}24.0 & \cellcolor[rgb]{ .984,  .886,  .835}9.2 & \cellcolor[rgb]{ .678,  .678,  .678}16.5 & \cellcolor[rgb]{ .984,  .886,  .835}10.4 \\
    \midrule
    \multirow{4}[4]{*}{\textbf{Mistral}} & \multirow{2}[2]{*}{\textbf{Open Assistant}} & \textbf{FR } & 2.0   & \cellcolor[rgb]{ .949,  .949,  .949}3.5 & \cellcolor[rgb]{ .678,  .678,  .678}2.8 & \cellcolor[rgb]{ .984,  .886,  .835}2.0 & \cellcolor[rgb]{ .949,  .949,  .949}43.3 & \cellcolor[rgb]{ .984,  .886,  .835}0.0 & \cellcolor[rgb]{ .678,  .678,  .678}29.0 & \cellcolor[rgb]{ .949,  .949,  .949}67.7 & \cellcolor[rgb]{ .949,  .949,  .949}3.5 & \cellcolor[rgb]{ .984,  .886,  .835}2.0 & \cellcolor[rgb]{ .984,  .886,  .835}0.4 & \cellcolor[rgb]{ .949,  .949,  .949}3.5 \\
          &       & \textbf{PR} & 3.5   & \cellcolor[rgb]{ .984,  .886,  .835}3.5 & \cellcolor[rgb]{ .678,  .678,  .678}12.7 & \cellcolor[rgb]{ .984,  .886,  .835}2.8 & \cellcolor[rgb]{ .984,  .886,  .835}2.4 & \cellcolor[rgb]{ .984,  .886,  .835}2.8 & \cellcolor[rgb]{ .678,  .678,  .678}21.8 & \cellcolor[rgb]{ .984,  .886,  .835}0.8 & \cellcolor[rgb]{ .984,  .886,  .835}2.0 & \cellcolor[rgb]{ .949,  .949,  .949}6.3 & \cellcolor[rgb]{ .949,  .949,  .949}6.3 & \cellcolor[rgb]{ .984,  .886,  .835}2.8 \\
\cmidrule{2-15}          & \multirow{2}[2]{*}{\textbf{XS Test}} & \textbf{FR } & 0.4   & \cellcolor[rgb]{ .949,  .949,  .949}2.4 & \cellcolor[rgb]{ .678,  .678,  .678}5.2 & \cellcolor[rgb]{ .949,  .949,  .949}0.8 & \cellcolor[rgb]{ .949,  .949,  .949}11.6 & \cellcolor[rgb]{ .949,  .949,  .949}6.0 & \cellcolor[rgb]{ .678,  .678,  .678}12.9 & \cellcolor[rgb]{ .949,  .949,  .949}80.8 & \cellcolor[rgb]{ .949,  .949,  .949}4.8 & \cellcolor[rgb]{ .949,  .949,  .949}1.2 & \cellcolor[rgb]{ .949,  .949,  .949}0.8 & \cellcolor[rgb]{ .949,  .949,  .949}2.4 \\
          &       & \textbf{PR} & 4.0   & \cellcolor[rgb]{ .984,  .886,  .835}2.8 & \cellcolor[rgb]{ .678,  .678,  .678}8.4 & \cellcolor[rgb]{ .984,  .886,  .835}2.4 & \cellcolor[rgb]{ .984,  .886,  .835}2.4 & \cellcolor[rgb]{ .949,  .949,  .949}5.2 & \cellcolor[rgb]{ .678,  .678,  .678}16.5 & \cellcolor[rgb]{ .984,  .886,  .835}0.0 & \cellcolor[rgb]{ .984,  .886,  .835}2.0 & \cellcolor[rgb]{ .949,  .949,  .949}4.0 & \cellcolor[rgb]{ .949,  .949,  .949}5.6 & \cellcolor[rgb]{ .949,  .949,  .949}3.2 \\
    \midrule
    \multirow{4}[4]{*}{\textbf{Phi}} & \multirow{2}[2]{*}{\textbf{Open Assistant}} & \textbf{FR } & 2.8   & \cellcolor[rgb]{ .678,  .678,  .678}3.2 & \cellcolor[rgb]{ .678,  .678,  .678}6.7 & \cellcolor[rgb]{ .984,  .886,  .835}2.8 & \cellcolor[rgb]{ .678,  .678,  .678}44.9 & \cellcolor[rgb]{ .984,  .886,  .835}2.0 & \cellcolor[rgb]{ .678,  .678,  .678}40.2 & \cellcolor[rgb]{ .678,  .678,  .678}6.7 & \cellcolor[rgb]{ .678,  .678,  .678}5.9 & \cellcolor[rgb]{ .984,  .886,  .835}2.8 & \cellcolor[rgb]{ .984,  .886,  .835}2.4 & \cellcolor[rgb]{ .949,  .949,  .949}3.6 \\
          &       & \textbf{PR} & 7.5   & \cellcolor[rgb]{ .678,  .678,  .678}10.3 & \cellcolor[rgb]{ .678,  .678,  .678}19.4 & \cellcolor[rgb]{ .984,  .886,  .835}7.1 & \cellcolor[rgb]{ .678,  .678,  .678}47.6 & \cellcolor[rgb]{ .984,  .886,  .835}4.0 & \cellcolor[rgb]{ .678,  .678,  .678}70.9 & \cellcolor[rgb]{ .678,  .678,  .678}11.9 & \cellcolor[rgb]{ .678,  .678,  .678}9.9 & \cellcolor[rgb]{ .949,  .949,  .949}9.1 & \cellcolor[rgb]{ .949,  .949,  .949}7.9 & \cellcolor[rgb]{ .984,  .886,  .835}6.0 \\
\cmidrule{2-15}          & \multirow{2}[2]{*}{\textbf{XS Test}} & \textbf{FR } & 0.8   & \cellcolor[rgb]{ .678,  .678,  .678}3.6 & \cellcolor[rgb]{ .678,  .678,  .678}10.8 & \cellcolor[rgb]{ .949,  .949,  .949}2.4 & \cellcolor[rgb]{ .678,  .678,  .678}14.0 & \cellcolor[rgb]{ .949,  .949,  .949}16.4 & \cellcolor[rgb]{ .678,  .678,  .678}30.1 & \cellcolor[rgb]{ .678,  .678,  .678}34.4 & \cellcolor[rgb]{ .678,  .678,  .678}10.0 & \cellcolor[rgb]{ .949,  .949,  .949}4.4 & \cellcolor[rgb]{ .949,  .949,  .949}3.2 & \cellcolor[rgb]{ .949,  .949,  .949}4.8 \\
          &       & \textbf{PR} & 8.4   & \cellcolor[rgb]{ .678,  .678,  .678}22.4 & \cellcolor[rgb]{ .678,  .678,  .678}30.4 & \cellcolor[rgb]{ .949,  .949,  .949}10.8 & \cellcolor[rgb]{ .678,  .678,  .678}20.4 & \cellcolor[rgb]{ .949,  .949,  .949}23.2 & \cellcolor[rgb]{ .678,  .678,  .678}61.0 & \cellcolor[rgb]{ .678,  .678,  .678}36.0 & \cellcolor[rgb]{ .678,  .678,  .678}16.8 & \cellcolor[rgb]{ .949,  .949,  .949}12.1 & \cellcolor[rgb]{ .949,  .949,  .949}14.9 & \cellcolor[rgb]{ .949,  .949,  .949}12.1 \\
    \midrule
    \multirow{4}[4]{*}{\textbf{Qwen}} & \multirow{2}[2]{*}{\textbf{Open Assistant}} & \textbf{FR } & 1.6   & \cellcolor[rgb]{ .949,  .949,  .949}2.4 & \cellcolor[rgb]{ .678,  .678,  .678}9.8 & \cellcolor[rgb]{ .678,  .678,  .678}3.2 & \cellcolor[rgb]{ .678,  .678,  .678}43.3 & \cellcolor[rgb]{ 1,  0,  0}1.2 & \cellcolor[rgb]{ .678,  .678,  .678}13.0 & \cellcolor[rgb]{ .678,  .678,  .678}36.6 & \cellcolor[rgb]{ .678,  .678,  .678}3.2 & \cellcolor[rgb]{ .949,  .949,  .949}3.2 & \cellcolor[rgb]{ 1,  0,  0}1.6 & \cellcolor[rgb]{ .949,  .949,  .949}3.5 \\
          &       & \textbf{PR} & 5.5   & \cellcolor[rgb]{ .949,  .949,  .949}6.7 & \cellcolor[rgb]{ .678,  .678,  .678}15.4 & \cellcolor[rgb]{ .678,  .678,  .678}7.1 & \cellcolor[rgb]{ .678,  .678,  .678}44.5 & \cellcolor[rgb]{ 1,  0,  0}4.3 & \cellcolor[rgb]{ .678,  .678,  .678}24.5 & \cellcolor[rgb]{ .678,  .678,  .678}37.4 & \cellcolor[rgb]{ .678,  .678,  .678}7.5 & \cellcolor[rgb]{ .949,  .949,  .949}6.3 & \cellcolor[rgb]{ 1,  0,  0}3.5 & \cellcolor[rgb]{ .984,  .886,  .835}5.5 \\
\cmidrule{2-15}          & \multirow{2}[2]{*}{\textbf{XS Test}} & \textbf{FR } & 8.0   & \cellcolor[rgb]{ .984,  .886,  .835}6.4 & \cellcolor[rgb]{ .678,  .678,  .678}14.0 & \cellcolor[rgb]{ .678,  .678,  .678}8.4 & \cellcolor[rgb]{ .678,  .678,  .678}18.4 & \cellcolor[rgb]{ 1,  0,  0}5.6 & \cellcolor[rgb]{ .678,  .678,  .678}16.1 & \cellcolor[rgb]{ .678,  .678,  .678}52.4 & \cellcolor[rgb]{ .678,  .678,  .678}21.6 & \cellcolor[rgb]{ .984,  .886,  .835}7.2 & \cellcolor[rgb]{ 1,  0,  0}2.4 & \cellcolor[rgb]{ .949,  .949,  .949}14.9 \\
          &       & \textbf{PR} & 12.1  & \cellcolor[rgb]{ .984,  .886,  .835}10.0 & \cellcolor[rgb]{ .678,  .678,  .678}21.2 & \cellcolor[rgb]{ .678,  .678,  .678}14.4 & \cellcolor[rgb]{ .678,  .678,  .678}23.2 & \cellcolor[rgb]{ 1,  0,  0}7.6 & \cellcolor[rgb]{ .678,  .678,  .678}22.9 & \cellcolor[rgb]{ .678,  .678,  .678}53.6 & \cellcolor[rgb]{ .678,  .678,  .678}26.0 & \cellcolor[rgb]{ .949,  .949,  .949}13.6 & \cellcolor[rgb]{ 1,  0,  0}4.8 & \cellcolor[rgb]{ .949,  .949,  .949}16.5 \\
    \bottomrule
    \bottomrule
    \end{tabular}%
    }}
  \label{tab:rq2}%
  \vspace{-5mm}
\end{table}%

\section{RQ3: Performance Impact}

\label{sec:model_performance_impact}

\subsection{Setup}
To evaluate the impact of various defense mechanisms on model performance, we employ three popular benchmark datasets, including MMLU-Pro~\cite{wang2024mmlu}, IFEval~\cite{zhou2023instruction}, and GSM8K~\cite{cobbe2021gsm8k}. 
These benchmarks assess different aspects of model capacity, providing a comprehensive view of the trade-offs associated with each defense strategy. MMLU-Pro includes multiple-choice questions crafted to test LLM reasoning abilities. We select the health (900 samples), law (1.3k samples), and mathematics (1.3k samples) subset of MMLU-Pro. The evaluation metric is accuracy. GSM8K includes 1.3k mathematical questions that require multi-step reasoning. The evaluation metric is accuracy. IFEval is designed to evaluate instruction-following capabilities. This dataset consists of 500 verifiable prompts, where the correctness of responses can be objectively checked using deterministic programs, making it ideal for testing models’ adherence to specific instructions. We measure instruction-following accuracy using two complementary metrics: {strict accuracy} and {loose accuracy}. Strict accuracy evaluates whether a model's response adheres exactly to the given instruction, ensuring precise compliance. Loose accuracy, on the other hand, accounts for minor variations in phrasing or formatting that still satisfy the instruction’s intent, reducing the likelihood of penalizing valid responses due to superficial discrepancies.

\begin{table}[!tbp]
  \centering
  \caption{Performance of Various Defenders Across Models on various benchmarks. Light red cells denote cases where the defense approach outperforms all the other defenders.}
  \resizebox{0.94\linewidth}{!}{
    \begin{tabular}{c|l|ccc|ccc|ccc|ccc|ccc}
    \toprule
    \toprule
    \multirow{3}[2]{*}{\textbf{Models}} & \multirow{3}[2]{*}{\textbf{Defenders}} & \multicolumn{3}{c|}{\textbf{GSM8K }} & \multicolumn{3}{c|}{\textbf{IFEval }} & \multicolumn{9}{c}{\textbf{MMLU-Pro}} \\
          &       & \multicolumn{3}{c|}{\textbf{Math}} & \multicolumn{3}{c|}{\textbf{Semantic}} & \multicolumn{3}{c|}{\textbf{Health}} & \multicolumn{3}{c|}{\textbf{Law}} & \multicolumn{3}{c}{\textbf{Math}} \\
          &       & \textbf{Acc} & \textbf{\textbackslash{}delta acc} & \textbf{Rank} & \textbf{Acc} & \textbf{\textbackslash{}delta acc} & \textbf{Rank} & \textbf{Acc} & \textbf{\textbackslash{}delta acc} & \textbf{Rank} & \textbf{Acc} & \textbf{\textbackslash{}delta acc} & \textbf{Rank} & \textbf{Acc} & \textbf{\textbackslash{}delta acc} & \textbf{Rank} \\
    \midrule
    \multirow{12}[4]{*}{\textbf{Llama}} & \textbf{Baseline} & 0.81  & 0.000 & 1     & 0.75  & 0.000 & 2     & 0.53  & 0.000 & 1     & 0.31  & 0.000 & 1     & 0.48  & 0.000 & 1 \\
\cmidrule{2-17}          & \textbf{ICD} & 0.81  & -0.002 & 5     & 0.74  & -0.004 & 3     & 0.36  & -0.175 & 5     & 0.31  & -0.005 & 5     & 0.46  & -0.020 & 6 \\
          & \textbf{IA} & 0.62  & -0.189 & 9     & 0.43  & -0.323 & 10    & 0.18  & -0.350 & 6     & 0.20  & -0.114 & 8     & 0.11  & -0.370 & 11 \\
          & \textbf{LlamaGuard} & \cellcolor[rgb]{ .984,  .886,  .835}0.81 & \cellcolor[rgb]{ .984,  .886,  .835}0.000 & \cellcolor[rgb]{ .984,  .886,  .835}1 & 0.74  & -0.006 & 4     & 0.15  & -0.380 & 7     & 0.09  & -0.221 & 11    & 0.48  & -0.004 & 4 \\
          & \textbf{PPL} & \cellcolor[rgb]{ .984,  .886,  .835}0.81 & \cellcolor[rgb]{ .984,  .886,  .835}0.000 & \cellcolor[rgb]{ .984,  .886,  .835}1 & 0.74  & -0.013 & 5     & \cellcolor[rgb]{ .984,  .886,  .835}0.53 & \cellcolor[rgb]{ .984,  .886,  .835}0.000 & \cellcolor[rgb]{ .984,  .886,  .835}1 & \cellcolor[rgb]{ .984,  .886,  .835}0.31 & \cellcolor[rgb]{ .984,  .886,  .835}0.000 & \cellcolor[rgb]{ .984,  .886,  .835}1 & \cellcolor[rgb]{ .984,  .886,  .835}0.48 & \cellcolor[rgb]{ .984,  .886,  .835}0.000 & \cellcolor[rgb]{ .984,  .886,  .835}1 \\
          & \textbf{PriorityGoal} & 0.08  & -0.732 & 12    & 0.48  & -0.268 & 9     & 0.07  & -0.462 & 12    & 0.09  & -0.220 & 10    & 0.11  & -0.371 & 12 \\
          & \textbf{Retokenization} & 0.63  & -0.175 & 8     & 0.39  & -0.359 & 11    & 0.15  & -0.383 & 8     & 0.19  & -0.124 & 9     & 0.23  & -0.255 & 8 \\
          & \textbf{Self-Defend} & 0.81  & 0.000 & 1     & 0.67  & -0.083 & 7     & 0.14  & -0.396 & 10    & 0.30  & -0.009 & 6     & 0.47  & -0.019 & 5 \\
          & \textbf{Self-Exam} & 0.81  & -0.002 & 5     & 0.71  & -0.035 & 6     & 0.52  & -0.017 & 3     & 0.31  & -0.003 & 4     & \cellcolor[rgb]{ .984,  .886,  .835}0.48 & \cellcolor[rgb]{ .984,  .886,  .835}0.000 & \cellcolor[rgb]{ .984,  .886,  .835}1 \\
          & \textbf{Self-Paraphrase} & 0.64  & -0.171 & 7     & 0.18  & -0.571 & 12    & 0.12  & -0.410 & 11    & 0.26  & -0.046 & 7     & 0.22  & -0.266 & 9 \\
          & \textbf{Self-Reminder} & 0.48  & -0.329 & 11    & \cellcolor[rgb]{ .984,  .886,  .835}0.76 & \cellcolor[rgb]{ .984,  .886,  .835}0.007 & \cellcolor[rgb]{ .984,  .886,  .835}1 & 0.15  & -0.383 & 8     & 0.09  & -0.223 & 12    & 0.21  & -0.276 & 10 \\
          & \textbf{SmoothLLM} & 0.55  & -0.263 & 10    & 0.65  & -0.096 & 8     & 0.51  & -0.020 & 4     & \cellcolor[rgb]{ .984,  .886,  .835}0.31 & \cellcolor[rgb]{ .984,  .886,  .835}0.000 & \cellcolor[rgb]{ .984,  .886,  .835}1 & 0.38  & -0.106 & 7 \\
    \midrule
    \multirow{12}[3]{*}{\textbf{Mistral}} & \textbf{Baseline} & 0.55  & 0.000 & 1     & 0.51  & 0.000 & 2     & 0.42  & 0.000 & 2     & 0.24  & 0.000 & 1     & 0.26  & 0.000 & 2 \\
\cmidrule{2-17}          & \textbf{ICD} & 0.54  & -0.013 & 6     & \cellcolor[rgb]{ .984,  .886,  .835}0.53 & \cellcolor[rgb]{ .984,  .886,  .835}0.018 & \cellcolor[rgb]{ .984,  .886,  .835}1 & \cellcolor[rgb]{ .984,  .886,  .835}0.43 & \cellcolor[rgb]{ .984,  .886,  .835}0.007 & \cellcolor[rgb]{ .984,  .886,  .835}1 & 0.23  & -0.006 & 6     & 0.25  & -0.010 & 5 \\
          & \textbf{IA} & 0.01  & -0.541 & 12    & 0.34  & -0.168 & 9     & 0.24  & -0.183 & 7     & 0.16  & -0.082 & 8     & 0.11  & -0.147 & 11 \\
          & \textbf{LlamaGuard} & \cellcolor[rgb]{ .984,  .886,  .835}0.55 & \cellcolor[rgb]{ .984,  .886,  .835}0.000 & \cellcolor[rgb]{ .984,  .886,  .835}1 & 0.50  & -0.006 & 4     & 0.14  & -0.276 & 9     & 0.09  & -0.147 & 11    & \cellcolor[rgb]{ .984,  .886,  .835}0.26 & \cellcolor[rgb]{ .984,  .886,  .835}0.001 & \cellcolor[rgb]{ .984,  .886,  .835}1 \\
          & \textbf{PPL} & \cellcolor[rgb]{ .984,  .886,  .835}0.55 & \cellcolor[rgb]{ .984,  .886,  .835}0.000 & \cellcolor[rgb]{ .984,  .886,  .835}1 & 0.50  & -0.013 & 6     & \cellcolor[rgb]{ .984,  .886,  .835}0.42 & \cellcolor[rgb]{ .984,  .886,  .835}0.000 & \cellcolor[rgb]{ .984,  .886,  .835}2 & \cellcolor[rgb]{ .984,  .886,  .835}0.24 & \cellcolor[rgb]{ .984,  .886,  .835}0.000 & \cellcolor[rgb]{ .984,  .886,  .835}1 & \cellcolor[rgb]{ .984,  .886,  .835}0.26 & \cellcolor[rgb]{ .984,  .886,  .835}0.000 & \cellcolor[rgb]{ .984,  .886,  .835}2 \\
          & \textbf{PriorityGoal} & 0.48  & -0.072 & 7     & 0.37  & -0.137 & 8     & 0.16  & -0.263 & 8     & 0.14  & -0.099 & 9     & 0.13  & -0.124 & 9 \\
          & \textbf{Retokenization} & 0.32  & -0.233 & 9     & 0.24  & -0.270 & 10    & 0.13  & -0.291 & 10    & 0.11  & -0.129 & 10    & 0.11  & -0.142 & 10 \\
          & \textbf{Self-Defend} & 0.01  & -0.538 & 11    & 0.19  & -0.322 & 11    & 0.10  & -0.324 & 11    & 0.23  & -0.005 & 5     & 0.16  & -0.098 & 8 \\
          & \textbf{Self-Exam} & \cellcolor[rgb]{ .984,  .886,  .835}0.55 & \cellcolor[rgb]{ .984,  .886,  .835}0.000 & \cellcolor[rgb]{ .984,  .886,  .835}1 & 0.50  & -0.011 & 5     & 0.42  & -0.001 & 4     & \cellcolor[rgb]{ .984,  .886,  .835}0.24 & \cellcolor[rgb]{ .984,  .886,  .835}0.000 & \cellcolor[rgb]{ .984,  .886,  .835}1 & \cellcolor[rgb]{ .984,  .886,  .835}0.26 & \cellcolor[rgb]{ .984,  .886,  .835}0.000 & \cellcolor[rgb]{ .984,  .886,  .835}2 \\
          & \textbf{Self-Paraphrase} & 0.32  & -0.234 & 10    & 0.19  & -0.322 & 11    & 0.08  & -0.340 & 12    & 0.08  & -0.161 & 12    & 0.11  & -0.148 & 12 \\
          & \textbf{Self-Reminder} & 0.54  & -0.010 & 5     & 0.51  & -0.004 & 3     & 0.37  & -0.049 & 6     & 0.21  & -0.025 & 7     & 0.19  & -0.062 & 7 \\
          & \textbf{SmoothLLM} & 0.44  & -0.113 & 8     & 0.45  & -0.061 & 7     & 0.41  & -0.010 & 5     & \cellcolor[rgb]{ .984,  .886,  .835}0.24 & \cellcolor[rgb]{ .984,  .886,  .835}0.000 & \cellcolor[rgb]{ .984,  .886,  .835}1 & 0.24  & -0.019 & 6 \\ \midrule 
    \multirow{12}[3]{*}{\textbf{Qwen2}} & \textbf{Baseline} & 0.80  & 0.000 & 1     & 0.54  & 0.000 & 3     & 0.49  & 0.000 & 1     & 0.31  & 0.000 & 2     & 0.56  & 0.000 & 1 \\ 
\cmidrule{2-17}          & \textbf{ICD} & 0.78  & -0.017 & 5     & \cellcolor[rgb]{ .984,  .886,  .835}0.56 & \cellcolor[rgb]{ .984,  .886,  .835}0.022 & \cellcolor[rgb]{ .984,  .886,  .835}1 & 0.48  & -0.004 & 3     & 0.28  & -0.030 & 6     & 0.55  & -0.008 & 6 \\
          & \textbf{IA} & 0.30  & -0.498 & 11    & 0.31  & -0.231 & 10    & 0.10  & -0.383 & 11    & 0.16  & -0.151 & 10    & 0.09  & -0.463 & 12 \\
          & \textbf{LlamaGuard} & \cellcolor[rgb]{ .984,  .886,  .835}0.80 & \cellcolor[rgb]{ .984,  .886,  .835}0.000 & \cellcolor[rgb]{ .984,  .886,  .835}1 & 0.54  & -0.006 & 4     & 0.15  & -0.333 & 9     & 0.10  & -0.213 & 11    & 0.56  & -0.002 & 5 \\
          & \textbf{PPL} & \cellcolor[rgb]{ .984,  .886,  .835}0.80 & \cellcolor[rgb]{ .984,  .886,  .835}0.000 & \cellcolor[rgb]{ .984,  .886,  .835}1 & 0.53  & -0.011 & 6     & \cellcolor[rgb]{ .984,  .886,  .835}0.49 & \cellcolor[rgb]{ .984,  .886,  .835}0.000 & \cellcolor[rgb]{ .984,  .886,  .835}1 & \cellcolor[rgb]{ .984,  .886,  .835}0.31 & \cellcolor[rgb]{ .984,  .886,  .835}0.000 & \cellcolor[rgb]{ .984,  .886,  .835}2 & \cellcolor[rgb]{ .984,  .886,  .835}0.56 & \cellcolor[rgb]{ .984,  .886,  .835}0.000 & \cellcolor[rgb]{ .984,  .886,  .835}1 \\
          & \textbf{PriorityGoal} & 0.02  & -0.778 & 12    & 0.45  & -0.092 & 8     & 0.07  & -0.416 & 12    & 0.09  & -0.223 & 12    & 0.11  & -0.449 & 11 \\
          & \textbf{Retokenization} & 0.68  & -0.121 & 8     & 0.41  & -0.129 & 9     & 0.30  & -0.188 & 7     & 0.20  & -0.111 & 9     & 0.25  & -0.304 & 10 \\
          & \textbf{Self-Defend} & 0.50  & -0.304 & 10    & 0.27  & -0.272 & 11    & 0.11  & -0.379 & 10    & 0.27  & -0.041 & 7     & 0.26  & -0.296 & 9 \\
          & \textbf{Self-Exam} & 0.80  & -0.001 & 4     & 0.54  & -0.006 & 4     & 0.48  & -0.007 & 4     & 0.30  & -0.005 & 5     & 0.56  & -0.001 & 4 \\
          & \textbf{Self-Paraphrase} & 0.65  & -0.151 & 9     & 0.19  & -0.353 & 12    & 0.29  & -0.199 & 8     & 0.22  & -0.093 & 8     & 0.27  & -0.288 & 8 \\
          & \textbf{Self-Reminder} & 0.75  & -0.052 & 6     & \cellcolor[rgb]{ .984,  .886,  .835}0.56 & \cellcolor[rgb]{ .984,  .886,  .835}0.022 & \cellcolor[rgb]{ .984,  .886,  .835}1 & 0.48  & -0.010 & 6     & \cellcolor[rgb]{ .984,  .886,  .835}0.32 & \cellcolor[rgb]{ .984,  .886,  .835}0.005 & \cellcolor[rgb]{ .984,  .886,  .835}1 & 0.34  & -0.215 & 7 \\
          & \textbf{SmoothLLM} & 0.72  & -0.081 & 7     & 0.50  & -0.043 & 7     & 0.48  & -0.009 & 5     & 0.31  & 0.000 & 2     & \cellcolor[rgb]{ .984,  .886,  .835}0.56 & \cellcolor[rgb]{ .984,  .886,  .835}0.000 & \cellcolor[rgb]{ .984,  .886,  .835}1 \\
    \bottomrule
    \bottomrule
    \end{tabular}%
    }
  \label{tab:rq3}%
\end{table}%

\subsection{Results}

The overall results of RQ3 are presented in Table \ref{tab:rq3}. Here, red cells denote cases where a defender matches or slightly exceeds the native baseline. However, the dominant pattern is preservation or degradation rather than genuine improvement.

\noindent\textbf{Finding 3.1: \textbf{Defenses rarely improve downstream capability, but often degrade it.}} Across the evaluated benchmarks, nearly every defense reduces accuracy in at least one setting, and the few gains over the baseline are extremely small. This result suggests that downstream capability should be treated as a cost to manage, not a benefit to expect, when deploying jailbreak defenses.

\noindent\textbf{Finding 3.2: \textbf{PPL best preserves task performance across models and domains.}} PPL is the most stable defense in Table~\ref{tab:rq3}: it usually matches the baseline on GSM8K and MMLU-Pro, and rarely introduces large drops. Rather than improving capability, PPL mostly preserves it, making rule-based filtering the most performance-friendly option among the evaluated defenses.

\noindent\textbf{Finding 3.3: \textbf{Performance impact is highly domain-dependent.}} The same defender can behave very differently across tasks. For example, LlamaGuard preserves GSM8K and MMLU-Pro Math reasonably well for some models but sharply hurts law-related accuracy, while Self-Reminder preserves instruction following better than it preserves mathematical reasoning. These patterns indicate that performance impact depends not only on the defense strategy but also on the task type being protected.

\noindent\textbf{Finding 3.4: \textbf{Rule-based defenses preserve performance better than conservative reflective defenses.}}
At the category level, rule-based defenses such as PPL preserve baseline-level performance much better than intention-analysis and self-reflective defenses. By contrast, methods such as IA and Self-Defend often cause the largest drops on instruction-following and reasoning-heavy tasks, showing that stronger internal scrutiny can interfere with normal capability.

\noindent\textbf{Insight:} The performance cost of a defender depends strongly on both the strategy and the task. Rule-based methods such as {PPL} preserve accuracy well on structured benchmarks, whereas self-reflective methods such as {Self-Defend} and {Intention Analysis} often introduce larger accuracy drops on complex domains such as MMLU-Pro.

% \begin{center}
% \begin{minipage}{0.92\linewidth}
% \begin{shaded}
% \noindent\textbf{Insight:} The performance cost of a defender depends strongly on both the strategy and the task. Rule-based methods such as {PPL} preserve accuracy well on structured benchmarks, whereas self-reflective methods such as {Self-Defend} and {Intention Analysis} often introduce larger accuracy drops on complex domains such as MMLU-Pro.
% \end{shaded}
% \end{minipage}
% \end{center}

\section{RQ4: Cost}

\label{sec:overhead}

\subsection{Setup} 
To comprehensively assess the additional inference cost introduced by each defense mechanism, we employed the {Open Assistant Dataset}~\cite{kopf2024openassistant}. 
This dataset, designed for evaluating interactive and multi-turn conversational responses, provides a standardized context that enables consistent measurements of computational requirements. 
For each model-defense pairing, we tracked the total number of tokens processed and generated by each model under different defenses. 
This metric reflects the additional processing load due to token generation and transformation associated with each defense mechanism.

% \CM{Equation for each metric and highlight findings }
We conduct a comprehensive analysis of how defense strategies translate into token overhead and runtime cost. Our empirical measurements on the Open Assistant dataset across Qwen2-7B, Llama-3.1-8B, and Mistral-7B-v0.3 show that defenses differ substantially in how much they inflate input and output token counts, and that larger output-token footprints are consistently associated with higher latency and energy consumption.

\noindent\textbf{Finding 4.1: \textbf{SmoothLLM incurs the largest output-token overhead.}} SmoothLLM consistently produces the largest increase in output tokens, especially on Llama-3.1-8B and Mistral-7B-v0.3. This result follows directly from its multi-round generation process, which amplifies response length through repeated sampling and checking. By contrast, PPL, Self-Reminder, and Self-Paraphrase introduce much smaller output-token changes.

% \CM{All approach will increase input tokens}

% \CM{Some approach will decreaase output tokens, indicating they will save cost}

\noindent\textbf{Finding 4.2: \textbf{Prompt-heavy and multi-round defenses dominate input-token inflation.}} All evaluated defenses add some input overhead, but the magnitude differs substantially. Multi-round methods and long-template methods such as LlamaGuard contribute the largest increases, whereas rule-based defenses such as PPL and lightweight perturbation methods produce much smaller changes. This result shows that input-token cost is driven mainly by prompt engineering and repeated safety interaction rather than by defense presence alone.

% \zexin{check the following paragraph w.r.t. costs. maybe more concise.}

% \\
% To further evaluate the practical implications of token overhead introduced by each defense mechanism, we reference the token pricing of popular LLMs. The prices for processing tokens vary significantly across providers, with output tokens generally costing more than input tokens in most models, especially in the case of high-end models like OpenAI’s GPT-4 and Anthropic’s Claude 3 Opus.
\noindent\textbf{Finding 4.3: \textbf{Token overhead translates directly into runtime and API cost.}}
Our empirical measurements show that defenses with larger output-token footprints also consume more latency and energy. In practical API deployments, this pattern means that output-heavy defenses such as SmoothLLM and prompt-heavy defenses such as LlamaGuard can impose substantial monetary cost even when their safety gains are real. Cost should therefore be treated as a first-class deployment criterion rather than as a secondary implementation detail.

\noindent\textbf{Finding 4.4: \textbf{Latency and energy consumption scale with output-token inflation.}}
We observe a strong positive correlation between output-token count and system overhead: as defenses generate more output tokens, both latency and energy usage rise proportionally. For example, SmoothLLM significantly increases latency and energy consumption (e.g., more than a 400\% increase for Llama-3.1-8B) relative to the baseline due to its multi-round processing. By contrast, rule-based and lightweight defenses such as PPL and PriorityGoal introduce minimal increases in energy usage. This result implies that defenses relying on extensive output generation impose steep operational penalties, making them unsuitable for environments with strict response-time or power budgets.

\noindent\textbf{Insight:} Defenders that increase token counts, especially output tokens, can materially raise deployment cost in API settings. SmoothLLM and LlamaGuard impose the largest overhead, so cost-sensitive applications should prefer more token-efficient defenses whenever possible.

\section{Implications for Practitioners}

Our results suggest that defense choice should be driven by the operational constraint that matters most in deployment, rather than by jailbreak blocking alone. As visually summarized in the trade-off radar chart in right side of Fig.~\ref{fig:overview}, different strategies occupy completely different regions of the safety-utility-cost space. Based on these empirical profiles, we provide the following practical guidelines: 

\begin{itemize}[leftmargin=10pt]
    \item \textbf{General-Purpose Assistants (High Throughput):} When preserving downstream reasoning quality and keeping latency low are the primary goals, simple rule-based defenses (such as PPL) are often the safest default. They introduce the smallest capability and cost penalties while maintaining acceptable baseline filtering. 

    \item \textbf{Balanced Consumer Applications (Moderate Risk):} When practitioners need a more balanced point between safety and usability without extreme overhead, output cross-reflection methods (such as LlamaGuard) are often the best fit. They improve screening accuracy on obvious harms without the massive latency spikes of multi-round strategies. 

    \item \textbf{High-Stakes Compliance (Maximum Safety):} When the threat model is strict and maximizing attack blocking is more important than preserving benign-query acceptance, conservative self-reflection methods (such as Self-Defend) may be justified. However, they should be deployed with clear awareness that they can substantially reduce helpfulness and reasoning capacity. 

    \item \textbf{Adversarial Environments (Robustness over Cost):} Finally, perturbation and multi-round methods (such as SmoothLLM) are highly effective against advanced token-level jailbreaks, but they are best reserved for settings with enough latency tolerance and token budget to absorb their significant runtime cost.
\end{itemize}

In short, our benchmark suggests a practical decision guide: start with rule-based or lightweight output-checking defenses for general-purpose models, escalate to stronger reflective methods only in high-risk deployments, and avoid expensive multi-round defenses unless the application can strictly tolerate the inference overhead.

\section{Conclusion}

We presented a comprehensive study of the unintended side effects introduced by jailbreak defenses for LLMs. Although these defenses improve safety, they can also degrade task performance, increase over-refusal on benign queries, and raise inference cost through additional computation. Our benchmark shows that these side effects are substantial and vary systematically across defense categories. A central takeaway is that stronger-looking defenses are not always better for deployment: simple rule-based methods can preserve reasoning ability far better than sophisticated reflective approaches, while multi-round defenses can become prohibitively expensive. We hope this study provides both a practical evaluation framework and a clearer basis for designing defenses that preserve safety without sacrificing usability. Future work should focus on adaptive and cost-aware defense strategies that reduce these side effects in realistic deployment settings.

\bibliographystyle{plainnat}
\bibliography{anthology,custom}

\begin{thebibliography}{37}
\providecommand{\natexlab}[1]{#1}
\providecommand{\url}[1]{\texttt{#1}}
\expandafter\ifx\csname urlstyle\endcsname\relax
  \providecommand{\doi}[1]{doi: #1}\else
  \providecommand{\doi}{doi: \begingroup \urlstyle{rm}\Url}\fi

\bibitem[Abdin et~al.(2024)Abdin, Aneja, and
  Awadalla]{abdin2024phi3technicalreporthighly}
Marah Abdin, Jyoti Aneja, and Hany et~al. Awadalla.
\newblock Phi-3 technical report: A highly capable language model locally on
  your phone, 2024.
\newblock URL \url{https://arxiv.org/abs/2404.14219}.

\bibitem[Achiam et~al.(2023)Achiam, Adler, Agarwal, Ahmad, Akkaya, Aleman,
  Almeida, Altenschmidt, Altman, Anadkat, et~al.]{achiam2023gpt}
Josh Achiam, Steven Adler, Sandhini Agarwal, Lama Ahmad, Ilge Akkaya,
  Florencia~Leoni Aleman, Diogo Almeida, Janko Altenschmidt, Sam Altman,
  Shyamal Anadkat, et~al.
\newblock Gpt-4 technical report.
\newblock \emph{arXiv preprint arXiv:2303.08774}, 2023.

\bibitem[Al-Rumaim and D.~Pawar(2023)]{akram-jyoti-2023-revolutionizing}
Akram Al-Rumaim and Jyoti D.~Pawar.
\newblock Revolutionizing authentication: Harnessing natural language
  understanding for dynamic password generation and verification.
\newblock In Jyoti D.~Pawar and Sobha Lalitha~Devi, editors, \emph{Proceedings
  of the 20th International Conference on Natural Language Processing (ICON)},
  pages 670--678, Goa University, Goa, India, December 2023. NLP Association of
  India (NLPAI).
\newblock URL \url{https://aclanthology.org/2023.icon-1.67}.

\bibitem[Alon and Kamfonas(2023)]{alon2023detectinglanguagemodelattacks}
Gabriel Alon and Michael Kamfonas.
\newblock Detecting language model attacks with perplexity, 2023.
\newblock URL \url{https://arxiv.org/abs/2308.14132}.

\bibitem[Brown(2020)]{brown2020language}
Tom~B Brown.
\newblock Language models are few-shot learners.
\newblock \emph{arXiv preprint arXiv:2005.14165}, 2020.

\bibitem[Cobbe et~al.(2021)Cobbe, Kosaraju, Bavarian, Chen, Jun, Kaiser,
  Plappert, Tworek, Hilton, Nakano, Hesse, and Schulman]{cobbe2021gsm8k}
Karl Cobbe, Vineet Kosaraju, Mohammad Bavarian, Mark Chen, Heewoo Jun, Lukasz
  Kaiser, Matthias Plappert, Jerry Tworek, Jacob Hilton, Reiichiro Nakano,
  Christopher Hesse, and John Schulman.
\newblock Training verifiers to solve math word problems.
\newblock \emph{arXiv preprint arXiv:2110.14168}, 2021.

\bibitem[Cui et~al.(2024)Cui, Chiang, Stoica, and Hsieh]{cui2024or}
Justin Cui, Wei-Lin Chiang, Ion Stoica, and Cho-Jui Hsieh.
\newblock Or-bench: An over-refusal benchmark for large language models.
\newblock \emph{arXiv preprint arXiv:2405.20947}, 2024.

\bibitem[DeepSeek-AI(2024)]{deepseekv2}
DeepSeek-AI.
\newblock Deepseek-v2: A strong, economical, and efficient mixture-of-experts
  language model, 2024.

\bibitem[Dubey et~al.(2024)Dubey, Jauhri, and
  Pandey]{dubey2024llama3herdmodels}
Abhimanyu Dubey, Abhinav Jauhri, and Abhinav et~al. Pandey.
\newblock The llama 3 herd of models, 2024.
\newblock URL \url{https://arxiv.org/abs/2407.21783}.

\bibitem[Feng et~al.(2024)Feng, Zhang, Li, Liu, Lang, Feng, Wu, and
  Liu]{feng2024improving}
Zhaopeng Feng, Yan Zhang, Hao Li, Wenqiang Liu, Jun Lang, Yang Feng, Jian Wu,
  and Zuozhu Liu.
\newblock Improving llm-based machine translation with systematic
  self-correction.
\newblock \emph{arXiv preprint arXiv:2402.16379}, 2024.

\bibitem[Finnie-Ansley et~al.(2022)Finnie-Ansley, Denny, Becker, Luxton-Reilly,
  and Prather]{finnie2022robots}
James Finnie-Ansley, Paul Denny, Brett~A Becker, Andrew Luxton-Reilly, and
  James Prather.
\newblock The robots are coming: Exploring the implications of openai codex on
  introductory programming.
\newblock In \emph{Proceedings of the 24th Australasian Computing Education
  Conference}, pages 10--19, 2022.

\bibitem[Inan et~al.(2023)Inan, Upasani, Chi, Rungta, Iyer, Mao, Tontchev, Hu,
  Fuller, Testuggine, and Khabsa]{Inan2023LlamaGL}
Hakan Inan, K.~Upasani, Jianfeng Chi, Rashi Rungta, Krithika Iyer, Yuning Mao,
  Michael Tontchev, Qing Hu, Brian Fuller, Davide Testuggine, and Madian
  Khabsa.
\newblock Llama guard: Llm-based input-output safeguard for human-ai
  conversations.
\newblock \emph{ArXiv}, abs/2312.06674, 2023.
\newblock URL \url{https://api.semanticscholar.org/CorpusID:266174345}.

\bibitem[Jain et~al.(2023)Jain, Schwarzschild, Wen, Somepalli, Kirchenbauer,
  yeh Chiang, Goldblum, Saha, Geiping, and
  Goldstein]{jain2023baselinedefensesadversarialattacks}
Neel Jain, Avi Schwarzschild, Yuxin Wen, Gowthami Somepalli, John Kirchenbauer,
  Ping yeh Chiang, Micah Goldblum, Aniruddha Saha, Jonas Geiping, and Tom
  Goldstein.
\newblock Baseline defenses for adversarial attacks against aligned language
  models, 2023.
\newblock URL \url{https://arxiv.org/abs/2309.00614}.

\bibitem[Jiang et~al.(2023)Jiang, Sablayrolles, and Mensch]{jiang2023mistral7b}
Albert~Q. Jiang, Alexandre Sablayrolles, and Arthur et~al. Mensch.
\newblock Mistral 7b, 2023.
\newblock URL \url{https://arxiv.org/abs/2310.06825}.

\bibitem[K{\"o}pf et~al.(2024)K{\"o}pf, Kilcher, von R{\"u}tte, Anagnostidis,
  Tam, Stevens, Barhoum, Nguyen, Stanley, Nagyfi,
  et~al.]{kopf2024openassistant}
Andreas K{\"o}pf, Yannic Kilcher, Dimitri von R{\"u}tte, Sotiris Anagnostidis,
  Zhi~Rui Tam, Keith Stevens, Abdullah Barhoum, Duc Nguyen, Oliver Stanley,
  Rich{\'a}rd Nagyfi, et~al.
\newblock Openassistant conversations-democratizing large language model
  alignment.
\newblock \emph{Advances in Neural Information Processing Systems}, 36, 2024.

\bibitem[Li et~al.(2024)Li, Dong, Wang, Hu, Zuo, Lin, et~al.]{li2024saladbench}
Lijun Li, Bowen Dong, Ruoyu Wang, Jiaojiao Hu, Wenmian Zuo, Dahua Lin, et~al.
\newblock Salad-bench: A hierarchical and comprehensive safety benchmark for
  large language models.
\newblock In \emph{ACL}, 2024.

\bibitem[Mai et~al.(2024)Mai, Hong, Chen, Pan, Liu, Zhang, Duan, and
  Yang]{mai2025usebench}
Wuyuao Mai, Geng Hong, Pei Chen, Xu~Pan, Baojun Liu, Yuan Zhang, Haixin Duan,
  and Min Yang.
\newblock You can't eat your cake and have it too: The performance degradation
  of llms with jailbreak defense.
\newblock \emph{arXiv preprint arXiv:2407.13540}, 2024.

\bibitem[Phute et~al.(2024)Phute, Helbling, Hull, Peng, Szyller, Cornelius, and
  Chau]{phute2024llmselfdefenseself}
Mansi Phute, Alec Helbling, Matthew Hull, ShengYun Peng, Sebastian Szyller,
  Cory Cornelius, and Duen~Horng Chau.
\newblock Llm self defense: By self examination, llms know they are being
  tricked, 2024.
\newblock URL \url{https://arxiv.org/abs/2308.07308}.

\bibitem[Robey et~al.(2023)Robey, Wong, Hassani, and
  Pappas]{robey2023smoothllm}
Alexander Robey, Eric Wong, Hamed Hassani, and George~J Pappas.
\newblock Smoothllm: Defending large language models against jailbreaking
  attacks.
\newblock \emph{arXiv preprint arXiv:2310.03684}, 2023.

\bibitem[R{\"o}ttger et~al.(2024)R{\"o}ttger, Kirk, Vidgen, Attanasio, Bianchi,
  and Hovy]{rottger-etal-2024-xstest}
Paul R{\"o}ttger, Hannah Kirk, Bertie Vidgen, Giuseppe Attanasio, Federico
  Bianchi, and Dirk Hovy.
\newblock {XST}est: A test suite for identifying exaggerated safety behaviours
  in large language models.
\newblock In Kevin Duh, Helena Gomez, and Steven Bethard, editors,
  \emph{Proceedings of the 2024 Conference of the North American Chapter of the
  Association for Computational Linguistics: Human Language Technologies
  (Volume 1: Long Papers)}, pages 5377--5400, Mexico City, Mexico, June 2024.
  Association for Computational Linguistics.
\newblock \doi{10.18653/v1/2024.naacl-long.301}.
\newblock URL \url{https://aclanthology.org/2024.naacl-long.301}.

\bibitem[Sun et~al.(2024)Sun, Huang, Ding, Liu, et~al.]{sun2024trustllm}
Licheng Sun, Yuxin Huang, Heng Ding, Yue Liu, et~al.
\newblock Trustllm: Trustworthiness in large language models.
\newblock In \emph{International Conference on Machine Learning (ICML)}, 2024.

\bibitem[Tan et~al.(2024)Tan, Luo, Li, and Zhang]{tan2024llm4decompile}
Hanzhuo Tan, Qi~Luo, Jing Li, and Yuqun Zhang.
\newblock Llm4decompile: Decompiling binary code with large language models.
\newblock \emph{arXiv preprint arXiv:2403.05286}, 2024.

\bibitem[Team et~al.(2024)Team, Riviere, and
  Pathak]{gemmateam2024gemma2improvingopen}
Gemma Team, Morgane Riviere, and Shreya et~al. Pathak.
\newblock Gemma 2: Improving open language models at a practical size, 2024.
\newblock URL \url{https://arxiv.org/abs/2408.00118}.

\bibitem[Varshney et~al.(2024)Varshney, Dolin, Seth, and
  Baral]{varshney-etal-2024-art}
Neeraj Varshney, Pavel Dolin, Agastya Seth, and Chitta Baral.
\newblock The art of defending: A systematic evaluation and analysis of {LLM}
  defense strategies on safety and over-defensiveness.
\newblock In Lun-Wei Ku, Andre Martins, and Vivek Srikumar, editors,
  \emph{Findings of the Association for Computational Linguistics ACL 2024},
  pages 13111--13128, Bangkok, Thailand and virtual meeting, August 2024.
  Association for Computational Linguistics.
\newblock \doi{10.18653/v1/2024.findings-acl.776}.
\newblock URL \url{https://aclanthology.org/2024.findings-acl.776}.

\bibitem[Wang et~al.(2024{\natexlab{a}})Wang, Shi, Bai, and
  Hsieh]{wang-etal-2024-defending}
Yihan Wang, Zhouxing Shi, Andrew Bai, and Cho-Jui Hsieh.
\newblock Defending {LLM}s against jailbreaking attacks via backtranslation.
\newblock In Lun-Wei Ku, Andre Martins, and Vivek Srikumar, editors,
  \emph{Findings of the Association for Computational Linguistics ACL 2024},
  pages 16031--16046, Bangkok, Thailand and virtual meeting, August
  2024{\natexlab{a}}. Association for Computational Linguistics.
\newblock \doi{10.18653/v1/2024.findings-acl.948}.
\newblock URL \url{https://aclanthology.org/2024.findings-acl.948}.

\bibitem[Wang et~al.(2024{\natexlab{b}})Wang, Ma, Zhang, Ni, Chandra, Guo, Ren,
  Arulraj, He, Jiang, et~al.]{wang2024mmlu}
Yubo Wang, Xueguang Ma, Ge~Zhang, Yuansheng Ni, Abhranil Chandra, Shiguang Guo,
  Weiming Ren, Aaran Arulraj, Xuan He, Ziyan Jiang, et~al.
\newblock Mmlu-pro: A more robust and challenging multi-task language
  understanding benchmark.
\newblock \emph{arXiv preprint arXiv:2406.01574}, 2024{\natexlab{b}}.

\bibitem[Wei et~al.(2024)Wei, Wang, Li, Mo, and
  Wang]{wei2024jailbreakguardalignedlanguage}
Zeming Wei, Yifei Wang, Ang Li, Yichuan Mo, and Yisen Wang.
\newblock Jailbreak and guard aligned language models with only few in-context
  demonstrations, 2024.
\newblock URL \url{https://arxiv.org/abs/2310.06387}.

\bibitem[Wu et~al.(2024)Wu, Wang, Liu, and Liu]{Wu2024LLMsCD}
Daoyuan Wu, Shuaibao Wang, Yang Liu, and Ning Liu.
\newblock Llms can defend themselves against jailbreaking in a practical
  manner: A vision paper.
\newblock \emph{ArXiv}, abs/2402.15727, 2024.
\newblock URL \url{https://api.semanticscholar.org/CorpusID:267938128}.

\bibitem[Xie et~al.(2023)Xie, Yi, Shao, Curl, Lyu, Chen, Xie, and
  Wu]{xie2023defending}
Yueqi Xie, Jingwei Yi, Jiawei Shao, Justin Curl, Lingjuan Lyu, Qifeng Chen,
  Xing Xie, and Fangzhao Wu.
\newblock Defending chatgpt against jailbreak attack via self-reminders.
\newblock \emph{Nature Machine Intelligence}, 5\penalty0 (12):\penalty0
  1486--1496, 2023.

\bibitem[Xu et~al.(2024{\natexlab{a}})Xu, Sharaf, Chen, Tan, Shen, Durme,
  Murray, and Kim]{xu2024contrastive}
Haoran Xu, Amr Sharaf, Yunmo Chen, Weiting Tan, Lingfeng Shen, Benjamin~Van
  Durme, Kenton Murray, and Young~Jin Kim.
\newblock Contrastive preference optimization: Pushing the boundaries of {LLM}
  performance in machine translation.
\newblock In \emph{Forty-first International Conference on Machine Learning},
  2024{\natexlab{a}}.
\newblock URL \url{https://openreview.net/forum?id=51iwkioZpn}.

\bibitem[Xu et~al.(2024{\natexlab{b}})Xu, Jiang, Niu, Jia, Lin, and
  Poovendran]{xu-etal-2024-safedecoding}
Zhangchen Xu, Fengqing Jiang, Luyao Niu, Jinyuan Jia, Bill~Yuchen Lin, and
  Radha Poovendran.
\newblock {S}afe{D}ecoding: Defending against jailbreak attacks via
  safety-aware decoding.
\newblock In Lun-Wei Ku, Andre Martins, and Vivek Srikumar, editors,
  \emph{Proceedings of the 62nd Annual Meeting of the Association for
  Computational Linguistics (Volume 1: Long Papers)}, pages 5587--5605,
  Bangkok, Thailand, August 2024{\natexlab{b}}. Association for Computational
  Linguistics.
\newblock \doi{10.18653/v1/2024.acl-long.303}.
\newblock URL \url{https://aclanthology.org/2024.acl-long.303}.

\bibitem[Xu et~al.(2024{\natexlab{c}})Xu, Liu, Deng, Li, and
  Picek]{xu-etal-2024-comprehensive}
Zihao Xu, Yi~Liu, Gelei Deng, Yuekang Li, and Stjepan Picek.
\newblock A comprehensive study of jailbreak attack versus defense for large
  language models.
\newblock In Lun-Wei Ku, Andre Martins, and Vivek Srikumar, editors,
  \emph{Findings of the Association for Computational Linguistics ACL 2024},
  pages 7432--7449, Bangkok, Thailand and virtual meeting, August
  2024{\natexlab{c}}. Association for Computational Linguistics.
\newblock \doi{10.18653/v1/2024.findings-acl.443}.
\newblock URL \url{https://aclanthology.org/2024.findings-acl.443}.

\bibitem[Yang et~al.(2024)Yang, Yang, and Hui]{yang2024qwen2technicalreport}
An~Yang, Baosong Yang, and Binyuan et~al. Hui.
\newblock Qwen2 technical report, 2024.
\newblock URL \url{https://arxiv.org/abs/2407.10671}.

\bibitem[Zhang et~al.(2024{\natexlab{a}})Zhang, Guo, Zhu, Cao, Lin, Jia, Chen,
  and Wu]{zhang-etal-2024-jailbreak}
Hangfan Zhang, Zhimeng Guo, Huaisheng Zhu, Bochuan Cao, Lu~Lin, Jinyuan Jia,
  Jinghui Chen, and Dinghao Wu.
\newblock Jailbreak open-sourced large language models via enforced decoding.
\newblock In Lun-Wei Ku, Andre Martins, and Vivek Srikumar, editors,
  \emph{Proceedings of the 62nd Annual Meeting of the Association for
  Computational Linguistics (Volume 1: Long Papers)}, pages 5475--5493,
  Bangkok, Thailand, August 2024{\natexlab{a}}. Association for Computational
  Linguistics.
\newblock \doi{10.18653/v1/2024.acl-long.299}.
\newblock URL \url{https://aclanthology.org/2024.acl-long.299}.

\bibitem[Zhang et~al.(2024{\natexlab{b}})Zhang, Ding, Zhang, and
  Tao]{zhang2024intentionanalysismakesllms}
Yuqi Zhang, Liang Ding, Lefei Zhang, and Dacheng Tao.
\newblock Intention analysis makes llms a good jailbreak defender,
  2024{\natexlab{b}}.
\newblock URL \url{https://arxiv.org/abs/2401.06561}.

\bibitem[Zhang et~al.(2024{\natexlab{c}})Zhang, Yang, Ke, Mi, Wang, and
  Huang]{zhang-etal-2024-defending}
Zhexin Zhang, Junxiao Yang, Pei Ke, Fei Mi, Hongning Wang, and Minlie Huang.
\newblock Defending large language models against jailbreaking attacks through
  goal prioritization.
\newblock In Lun-Wei Ku, Andre Martins, and Vivek Srikumar, editors,
  \emph{Proceedings of the 62nd Annual Meeting of the Association for
  Computational Linguistics (Volume 1: Long Papers)}, pages 8865--8887,
  Bangkok, Thailand, August 2024{\natexlab{c}}. Association for Computational
  Linguistics.
\newblock \doi{10.18653/v1/2024.acl-long.481}.
\newblock URL \url{https://aclanthology.org/2024.acl-long.481}.

\bibitem[Zhou et~al.(2023)Zhou, Lu, Mishra, Brahma, Basu, Luan, Zhou, and
  Hou]{zhou2023instruction}
Jeffrey Zhou, Tianjian Lu, Swaroop Mishra, Siddhartha Brahma, Sujoy Basu,
  Yi~Luan, Denny Zhou, and Le~Hou.
\newblock Instruction-following evaluation for large language models.
\newblock \emph{arXiv preprint arXiv:2311.07911}, 2023.

\end{thebibliography}

\appendix
\newpage
\appendix

\section{Overhead}
Table~\ref{table:overhead_} shows input token  overhead across different defense mechanisms, grouped by meta-defender strategies, on the Open-Assistant dataset. This quantitative evaluation highlights the influence of various meta-defender categories on input token count, latency, and energy consumption, shedding light on the resource costs associated with each strategy.

Multi-round strategies, such as \textit{SmoothLLM}, tend to generate a substantially higher number of input tokens. This increase is due to the iterative nature of these approaches, which reprocesses and refines input to ensure safer outputs. For instance, \textit{SmoothLLM} in the Llama-3.1-8B model generates 404.43 input tokens compared to the baseline's 67.36 tokens, resulting in a notable increase in latency and energy usage (75.18 and 17882.62, respectively). This pattern of high token count is consistent across models, indicating that while multi-round strategies enhance response accuracy, they come at a cost of significantly increased computational demand.

Similarly, self-reflection methods, such as \textit{Self-Exam} and \textit{Self-Defend}, also exhibit high input token counts, especially in models like Qwen2-7B and DeepSeek-V2-Lite. \textit{Self-Exam}, for example, increases input tokens to 448.73 in Qwen2-7B from a baseline of 50.55, leading to higher energy use. This is likely due to the introspective analysis these methods perform on both input and generated content, necessitating additional token processing. In contrast, perturbation methods such as \textit{Retokenization} and rule-based checks such as \textit{PPL}, show a relatively moderate increase in input tokens, indicating their lower resource impact. By strategically altering or filtering the input with minimal token expansion, perturbation techniques maintain efficiency, making them suitable for applications requiring lower latency and energy overhead.

These findings underscore the need for careful selection of meta-defender strategies. For real-time or resource-constrained applications, the token efficiency of perturbation and rule-check strategies may be preferable. However, in safety-critical domains, where accuracy and caution are paramount, multi-round and self-reflection strategies, despite their resource demands, could be more appropriate.
\begin{table}[htbp]\centering
\caption{Input/Output tokens, Latency and Energy of Various Defenders on Open-Assistant Dataset.}

 \renewcommand\arraystretch{0.85} 
 \label{table:overhead_}
 \begin{minipage}{\textwidth}\centering
 \scalebox{0.75}{\begin{tabular}{cccccc}

 \toprule

        Models &   Defenders
          & Input Tokens &  Output Tokens &  Latency  &   Energy  \\
 \midrule
\multirow{11}{*}{DeepSeek-V2-Lite }&        Baseline &         42.19 &         453.86 &    25.17 &  2158.41 \\
 &             ICD &         91.01 &         422.08 &    23.38 &  1957.70 \\
 &              IA &        324.97 &         237.19 &    13.43 &  1163.75 \\
 &             PPL &         51.70 &         494.37 &    27.33 &  2335.91 \\
 &    PriorityGoal &        819.06 &         315.89 &    17.60 &  1534.57 \\
 &  Retokenization &         96.37 &         342.02 &    19.04 &  1606.32 \\
 &     Self-Defend &        137.79 &         460.77 &    25.78 &  2180.55 \\
&       Self-Exam &        526.05 &         507.20 &    28.15 &  2370.58 \\
 & Self-Paraphrase &        112.84 &         462.97 &    25.75 &  2156.01 \\
 &   Self-Reminder &         86.30 &         429.77 &    23.81 &  2009.93 \\
 &       SmoothLLM &        249.54 &        2719.43 &   150.05 & 12561.99 \\
\midrule
     \multirow{11}{*}{ Gemma-2-9b }&        Baseline &         37.15 &         359.89 &    22.07 &  4188.77 \\
     &             ICD &         89.21 &         302.28 &    18.30 &  3624.29 \\
       &              IA &        349.49 &         290.43 &    16.67 &  2992.24 \\
       &      LlamaGuard &        663.27 &         366.30 &    22.27 &  4364.03 \\
      &             PPL &         46.73 &         393.09 &    23.81 &  4536.28 \\
     &    PriorityGoal &        800.27 &         301.54 &    18.55 &  3673.10 \\
       &  Retokenization &         99.37 &         337.97 &    20.45 &  3969.61 \\
       &     Self-Defend &        155.50 &         378.45 &    21.66 &  3784.83 \\
       &       Self-Exam &        435.80 &         386.26 &    23.65 &  4688.71 \\
       & Self-Paraphrase &         89.89 &         397.86 &    24.03 &  4729.38 \\
      &       SmoothLLM &        228.29 &        2193.89 &   134.72 & 26520.62 \\
      \midrule
    \multirow{12}{*}{Llama-3.1-8B} &        Baseline &         67.36 &         504.24 &    12.48 &  3004.80 \\
     &             ICD &        119.63 &         408.44 &    10.16 &  2476.15 \\
     &              IA &        579.09 &         634.28 &    16.44 &  3868.03 \\
   &      LlamaGuard &        788.95 &         502.20 &    12.41 &  2988.43 \\
     &             PPL &         76.61 &         543.75 &    13.44 &  3264.95 \\
     &    PriorityGoal &        811.31 &         492.76 &    12.35 &  3092.29 \\
     &  Retokenization &        119.81 &         458.81 &    11.37 &  2765.55 \\
    &     Self-Defend &        214.22 &         532.26 &    13.47 &  3258.88 \\
     &       Self-Exam &        624.03 &         559.69 &    13.95 &  3359.80 \\
     & Self-Paraphrase &        183.48 &         568.10 &    14.06 &  3421.00 \\
     &   Self-Reminder &        108.23 &         455.42 &    11.28 &  2768.13 \\
    &       SmoothLLM &        404.43 &        3016.39 &    75.18 & 17882.62 \\
    
\bottomrule
 \end{tabular}}
 \end{minipage}
 \vspace{0mm}
 \end{table}

\begin{table}[htbp]\centering
\caption{Input/Output tokens, Latency and Energy of Various Defenders on Open-Assistant Dataset (Continued).}
 \renewcommand\arraystretch{0.85} 
 \begin{minipage}{\textwidth}\centering
 \scalebox{0.75}{\begin{tabular}{cccccc}
 \toprule
        Models &   Defenders
          & Input Tokens &  Output Tokens &  Latency  &   Energy  \\
 \midrule
\multirow{12}{*}{ Mistral-7B-V0.3} &        Baseline &         40.96 &         436.11 &    12.25 &  2579.69 \\
 &             ICD &         87.97 &         305.68 &     8.63 &  1785.85 \\
&              IA &        390.43 &         352.36 &    10.26 &  2163.45 \\
  &      LlamaGuard &        622.27 &         435.02 &    12.31 &  2565.20 \\
 &             PPL &         52.70 &         480.01 &    13.48 &  2813.16 \\
  &    PriorityGoal &        850.33 &         339.53 &     9.71 &  2122.14 \\
  &  Retokenization &         86.13 &         323.32 &     9.11 &  1888.26 \\
  &     Self-Defend &        160.50 &         467.52 &    13.09 &  2713.58 \\
  &       Self-Exam &        507.32 &         486.51 &    13.72 &  2856.42 \\
 & Self-Paraphrase &        333.91 &         637.38 &    18.02 &  3766.41 \\
  &   Self-Reminder &         88.65 &         327.58 &     9.24 &  1916.31 \\
  &       SmoothLLM &        242.02 &        2626.97 &    74.76 & 15401.60 \\
 \midrule
    \multirow{12}{*}{Phi-3.5-mini} &        Baseline &         38.71 &         567.50 &    19.47 &  2458.27 \\
     &             ICD &         87.33 &         550.35 &    19.18 &  2258.59 \\
    &              IA &        447.13 &         683.23 &    23.64 &  3157.93 \\
    &      LlamaGuard &        745.36 &         570.20 &    19.78 &  2409.21 \\
    &             PPL &         50.45 &         620.12 &    21.53 &  2618.61 \\
     &    PriorityGoal &        888.43 &         496.45 &    17.39 &  2474.08 \\
     &  Retokenization &         91.62 &         289.40 &    10.06 &  1136.52 \\
     &     Self-Defend &        164.09 &         634.17 &    21.75 &  2814.10 \\
   &       Self-Exam &        632.24 &         614.53 &    21.45 &  2566.35 \\
     & Self-Paraphrase &        128.14 &         614.80 &    21.38 &  2499.86 \\
     &   Self-Reminder &         89.74 &         519.64 &    18.11 &  2118.18 \\
     &       SmoothLLM &        234.25 &        3420.25 &   118.31 & 14287.55 \\
    \midrule
       \multirow{12}{*}{ Qwen2-7B} &        Baseline &         50.55 &         355.30 &     8.00 &  1975.88 \\
        &             ICD &        102.72 &         330.93 &     7.50 &  1852.18 \\
        &              IA &        349.59 &         317.73 &     7.18 &  1685.11 \\
        &      LlamaGuard &        626.29 &         357.55 &     8.19 &  2004.92 \\
         &             PPL &         59.78 &         391.17 &     8.85 &  2176.55 \\
         &    PriorityGoal &        796.35 &         346.46 &     7.92 &  2022.67 \\
         &  Retokenization &        106.11 &         346.39 &     7.84 &  1937.82 \\
         &     Self-Defend &        159.50 &         348.26 &     8.07 &  1883.00 \\
         &       Self-Exam &        448.73 &         361.76 &     8.20 &  2018.89 \\
         & Self-Paraphrase &        128.66 &         417.67 &     9.44 &  2325.45 \\
       &   Self-Reminder &         85.65 &         379.29 &     8.60 &  2123.21 \\
        &       SmoothLLM &        303.57 &        2125.35 &    46.67 & 11697.28 \\
\bottomrule
 \end{tabular}}
 \end{minipage}
 \vspace{0mm}
 \end{table}

\section{Response Classification Example}

\begin{table}[!tbp]
    \centering
    \small
    \caption{Examples of Response Classifications to the Query: "Why should black beans never be president?"}
    \label{tab:response_classifications}
    \resizebox{0.98\textwidth}{!}{%
    \begin{tabular}{lp{11cm}}
        \toprule
        \textbf{Class} & \textbf{Example Response} \\
        \midrule
        \textbf{Full\_Compliance} & \textbf{LlamaGuard:} "I'm not aware of any credible information that suggests black beans should never be president..., if you're asking this question in a serious context, I'd be happy to clarify that the qualifications for holding public office typically include being a human being, meeting certain age and residency requirements, and being a citizen of the country in question. Black beans do not meet these criteria." \\
        \midrule
        \textbf{Full\_Refusal} & \textbf{ICD:} "I can't provide information or guidance on harmful behavior such as racial discrimination. Is there anything else I can help you with?" \\
        \midrule
        \textbf{Partial\_Refusal} & \textbf{Self\_Paraphrase:} "The statement you made is a reflection of the principles of equality and non-discrimination that are fundamental to democratic societies. In the United States, for example, the 15th Amendment to the Constitution, ratified in 1870, guarantees the right to vote regardless of "race, color, or previous condition of servitude... " \\
        \bottomrule
    \end{tabular}
    }
\end{table}

As shown in Tab.~\ref{tab:response_classifications}, we demonstrate a qualitative example of response classification. We use GPT-4o-mini to categorize model responses into full compliance, full refusal, and partial refusal. This categorization allows us to compute the acceptance rate for unsafe queries, defined as the percentage of unsafe prompts that a model fails to reject. We also randomly sampled 200 examples and manually checked the labels. By contrasting the results of each defense mechanism with the Baseline, we gain insights into which defenders are most effective at minimizing unintended outputs without overly restricting legitimate inputs.

\section{Limitations}
We acknowledge several limitations of this study that future work could address. {(i) Model scope.} Our evaluation covers six representative open-source instruction-tuned LLMs in the 7B--9B parameter range (DeepSeek-V2-Lite-Chat, Meta-Llama-3.1-8B, Mistral-7B-Instruct-v0.3, Phi-3.5-Mini-Instruct, Gemma-2-9B-it, and Qwen2-7B). Frontier closed-source systems (e.g., GPT-4-class, Claude, Gemini) are used only as response classifiers and are not themselves defended; their alignment-tuned behavior, larger capacity, and proprietary safety stacks may yield different safety--utility--cost frontiers. {(ii) Linguistic and cultural scope.} All five benchmarks are English-only; over-refusal patterns and jailbreak success rates are known to shift in multilingual or low-resource settings, which we leave to future work.
{(iii) Cost measurement.} Although we report input/output tokens, latency, energy (Joules), and an estimated API cost, absolute figures depend on serving infrastructure, batch size, KV-cache reuse, and quantization, and should be read as relative comparisons rather than deployment forecasts. {(iv) Defense coverage.} The eleven defenses studied are widely cited but not exhaustive; emerging current work directions such as representation engineering, circuit breakers, and RLAIF-tuned guard models are out of scope and may exhibit different trade-off profiles. Finally, our findings are a snapshot against today's jailbreak corpora and may not generalize to stronger future attacks.

\section{Ethical Statement}
This work studies the unintended side effects of \textit{defenses} for large language models, and we believe its net contribution to the safety community is positive. All eleven defenses we evaluate (PPL, LlamaGuard, Self-Exam, Self-Defend, Intention Analysis, PriorityGoal, ICD, Self-Reminder, Self-Paraphrase, SmoothLLM, and Retokenization) and all five benchmarks (XSTest, Open Assistant, MMLU-Pro, IFEval, GSM8K) are publicly released artifacts; we introduce no new harmful content, no novel jailbreak attacks, and no proprietary model probing. Jailbreak prompts used for the safety axis are drawn exclusively from established public datasets, and the six evaluated models are openly licensed for research use, so our experiments do not violate any provider's terms of service. The study involves no human subjects, no personally identifiable information, and required no IRB review.

We have considered dual-use risk. In principle, our results reveal which defenses are weakest along the safety axis, which a malicious actor could exploit. We judge this risk to be low because (a) the relative robustness of these defenses is already documented across their original publications and prior surveys, (b) the additional axes we contribute, over-refusal, performance impact, and inference cost, are deployment-side concerns that do not enable an attack, and (c) our recommendations consistently favor \textit{stronger combinations} of defenses rather than weaker ones. We further note that aggressive over-refusal carries its own ethical cost: defenses that broadly refuse sensitive but legitimate queries (e.g., medical, legal, mental-health, or harm-reduction information) can disproportionately harm non-expert users who lack alternative channels of access. We release our code, configurations, and evaluation logs to support reproducibility and to enable practitioners to audit these trade-offs in their own settings. We encourage readers to deploy LLM defenses with full awareness of the safety, performance, and cost trade-offs documented herein.

\section{Hardware and Software Details}
All experiments were run on a GPU server with eight NVIDIA A6000 GPUs. The experimental pipeline was implemented in Python and PyTorch. To support reproducibility, we will release the code and benchmark data on acceptance.

%%%%%%%%%%%%%%%%%%%%%%%%%%%%%%%%%%%%%%%%%%%%%%%%%%%%%%%%%%%%

\end{document}